\documentclass[conference]{IEEEtran}
\usepackage{amssymb}
\renewcommand\IEEEkeywordsname{Keywords}
\IEEEoverridecommandlockouts
\usepackage{graphicx}
\usepackage[caption=false,font=footnotesize]{subfig}
\usepackage{cite}
\usepackage{picinpar}
\usepackage{amsmath}
\usepackage{url}
\usepackage{flushend}
\usepackage[latin1]{inputenc}
\usepackage{colortbl}
\usepackage{pifont}
\usepackage{color}
\usepackage{alltt}
\usepackage[hidelinks]{hyperref}
\usepackage{enumerate}
\usepackage{siunitx}
\usepackage{pbox}
\usepackage{amsfonts}
\usepackage{bm}
\usepackage[export]{adjustbox}
\usepackage{empheq}
\usepackage{xparse,mathtools}

\newtheorem{remark}{Remark}
\newcommand{\email}[1]{{\urlstyle{tt}\nolinkurl{#1}}}

\title{\LARGE\bf
Automated Sailboat Navigation using Controllability-Aware Nonlinear Model Predictive Control under Stochastic Winds
}
\author{Junzhuo~Wu$^1$,
        \text{Ya-Jun~Pan$^1$},~\IEEEmembership{Senior~Member,~IEEE,}
        \text{Chao~Shen$^2$},~\IEEEmembership{Member,~IEEE,}\\
        \text{Sean~Smith$^3$},~\IEEEmembership{Student~Member,~IEEE,}
        \text{and~Emmanuel~Witrant$^3$}
\thanks{This work is supported by the Natural Sciences and Engineering Research Council (NSERC) of Canada and the French Agence Nationale de la Recherche (ANR) AutoSail project (ANR-25-CE23-6846). For the purpose of open access, the author has applied a CC BY public copyright licence to any Author Accepted Manuscript (AAM) version arising from this submission.}
\thanks{$^1$J. Wu and Y.J. Pan are with the Advanced Control and Mechatronics Lab, Department of Mechanical Engineering, Dalhousie University, Halifax, Canada, B3H 4R2 (\email{jn574392@dal.ca}, \email{yajun.pan@dal.ca}). $^2$C. Shen is with the Department of Systems and Computer Engineering, Carleton University, Ottawa, K1S-5B6, Canada (\email{shenchao@sce.carleton.ca}). $^3$S. Smith and E. Witrant are with the GIPSA-lab, Universit\'e Grenoble Alpes - CNRS, F-38000, Grenoble, France, and the Department of Mechanical Engineering, Dalhousie University, Halifax B3H 4R2, Nova Scotia, Canada (\email{s.smith@dal.ca}, \email{emmanuel.witrant@univ-grenoble-alpes.fr}).}
}

\begin{document}
\maketitle
\begin{abstract}
This paper presents a systematic approach to plan and execute a time-efficient sailboat trajectory under the challenging stochastic wind conditions for automated sailboat. Unlike deterministic scenarios, stochastic wind  disturbances introduce challenges such as gust unpredictability, directional shifts, and fluctuating apparent wind speeds. These characteristics render traditional planning methods, such as Line-of-Sight (LoS) and waypoint approaches, to be ineffective and not applicable, as they often rely on static or deterministic environmental models. This paper proposed a new path planner and controller based on nonlinear model predictive control method. It is compared with a  baseline planner and controller. A Lie-algebraic analysis of the sailboat dynamics is carried out to identify the operating conditions under which the vessel loses first-order control authority in surge, and these conditions are embedded as constraints in the NMPC. Finally, simulation results are provided to validate the proposed framework.
\end{abstract}

\section{Introduction}

Autonomous sailboat navigation in marine environments is fundamentally challenged by the stochastic nature of wind. Unlike deterministic assumptions commonly adopted in classical planning and guidance strategies, wind fields exhibit significant temporal variability, including unpredictable gusts, directional shifts, and fluctuations in apparent wind speed. These uncertainties directly affect sail-generated propulsion and complicate time-optimal trajectory planning, often causing degraded performance or infeasibility under stochastic wind conditions.

To address wind uncertainty, several studies formulate sailboat trajectory planning as a stochastic optimal control problem. In \cite{Ferretti}, an optimal route planning framework was proposed using dynamic programming, where the sailboat is modeled as a point mass and the objective is to minimize travel time over a randomly evolving wind field. Building upon this work, \cite{sto} introduced a semi-Lagrangian dynamic programming approach with probabilistic wind modeling and an explicit tack-switching operator, significantly reducing computational complexity. While these methods provide insight into wind-aware planning, their reliance on point mass models neglects nonlinear dynamics, actuation limits, and maneuvering constraints, which may lead to trajectories that are not feasible for real sailboats.

Besides the stochastic nature of wind, automated sailboat control remains complex due to strong nonlinearities and coupling between states and control inputs. Nonlinear Model Predictive Control (NMPC) is well suited for this problem as it can handle multivariable nonlinear systems and enforce constraints directly within the control structure. Although only limited MPC-based sailboat control works have been proposed \cite{jmse11020460,TIPSUWAN2023114879}, NMPC has been widely applied in other marine domains, including autonomous marine vehicles, demonstrating robustness to environmental disturbances and modeling uncertainties \cite{shen,Abdelaal,8126875,10640017,7799190,9816891,shen2023marinebook}. Motivated by these developments, this paper adopts the stochastic planning framework of \cite{sto} as a benchmark and introduces an NMPC-based planner and controller for time-efficient autonomous sailboat navigation under stochastic wind conditions. 

The main contributions are: (i) a new integration of stochastic wind-aware trajectory planning with a realistic MPC-based control architecture, (ii) a Lie-algebraic analysis of the control authority of the sailboat that identifies maneuvering-degraded operating conditions and translates them into explicit NMPC constraints, and (iii) simulation results under deterministic and stochastic winds showing improved feasibility and robustness compared with an NMPC baseline without these constraints.


\section{Problem Formulation}
\subsection{Baseline Planner}
\begin{figure}[htbp]
	\centering
	\subfloat{\includegraphics[width=0.45\linewidth]{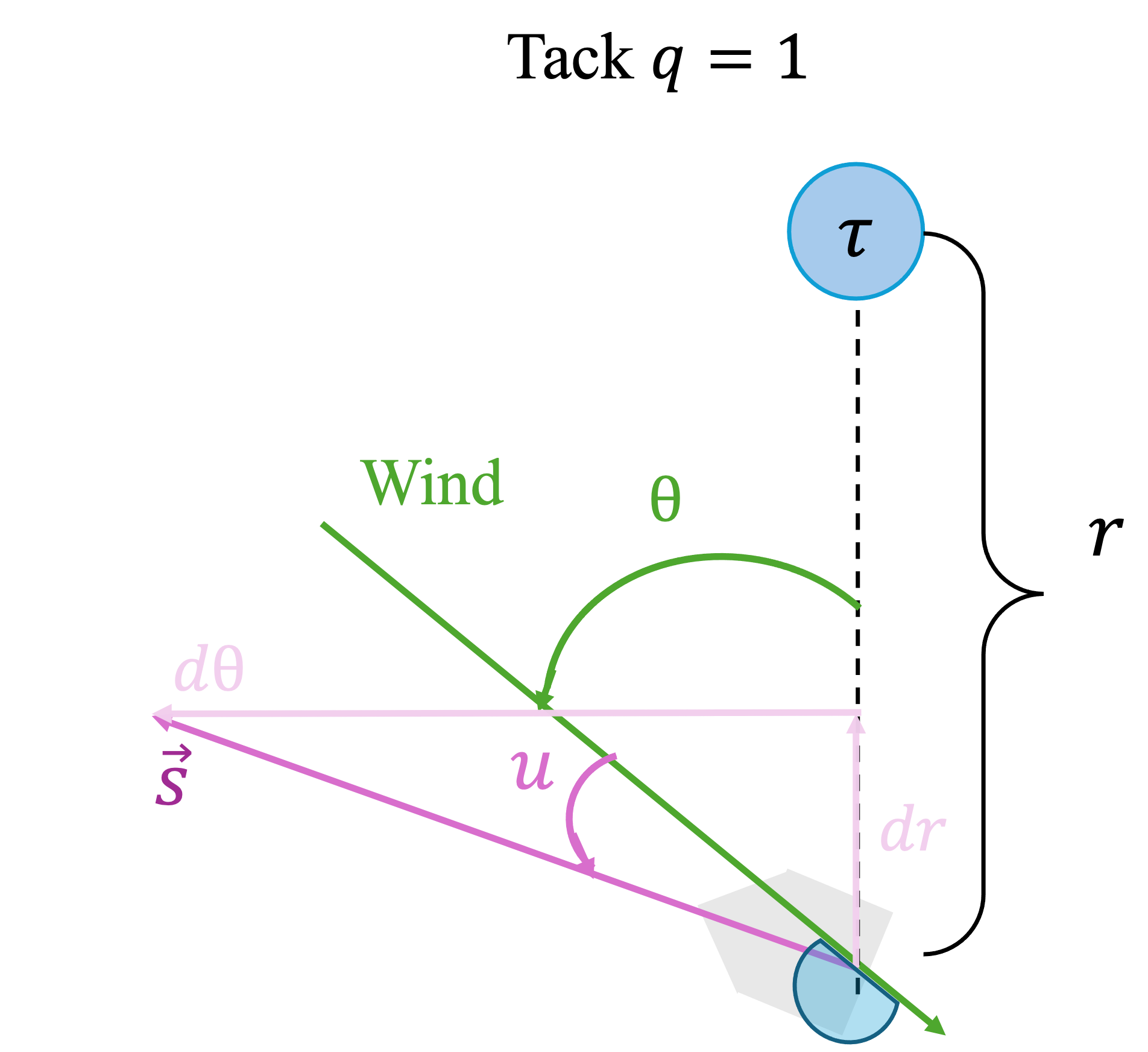}}
	\subfloat{\includegraphics[width=0.45\linewidth]{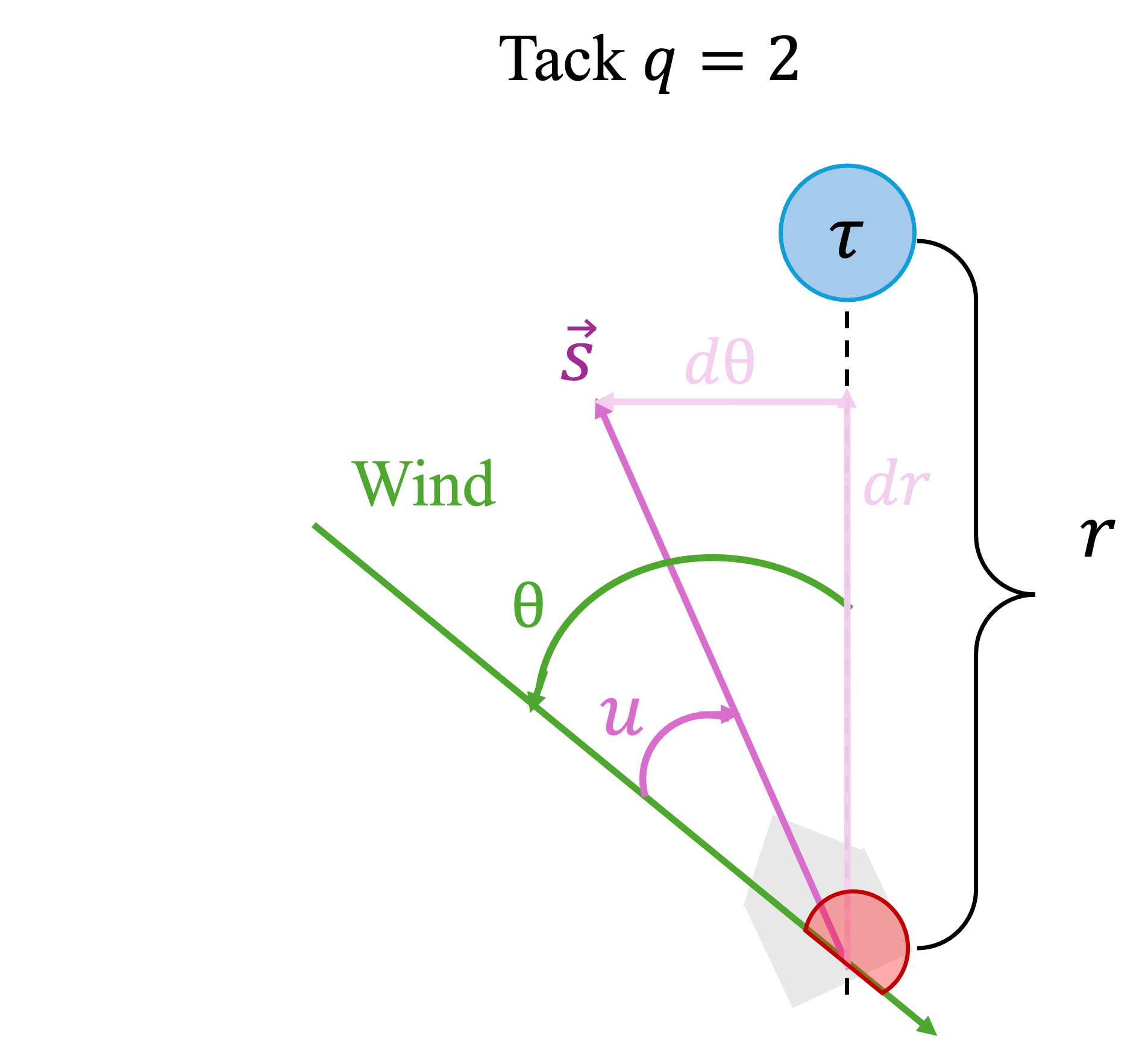}}
\caption{Stochastic path planning setup for sailboats modeled as point mass, with $q=1,2$.}\label{c5_base}
\end{figure}

\begin{figure}[ht]
	\includegraphics[width=.45\linewidth,center]{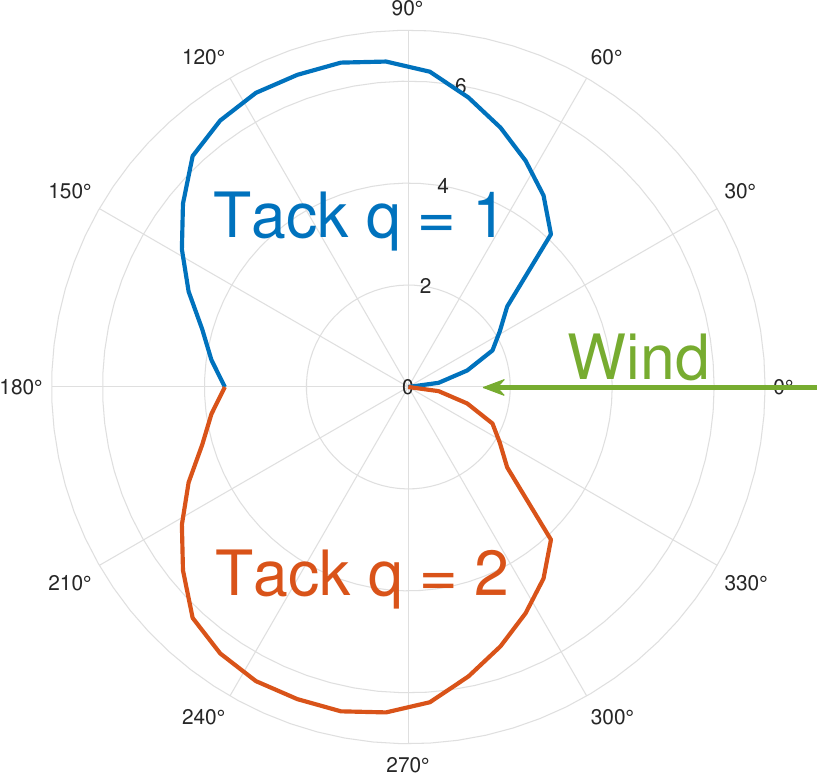}
	\caption{Polar speed plot.}
	\label{c5_polar} 
\end{figure}

Following the stochastic optimal control framework in \cite{sto}, shown in Fig.~\ref{c5_base}, the automated sailboat is modeled as a point mass subject to stochastic wind disturbances. 
The wind direction is treated as a stochastic process, while the wind speed is assumed deterministic. A discrete tack variable $q\in\{1,2\}$ indicates the wind side relative to the hull:$q=1$ and $q=2$ correspond to the wind on the starboard and port sides, respectively. The resulting stochastic dynamics are shown below:
\begin{equation}
\left\{
\begin{aligned}
dr &= -s(u)\cos\!\left(\theta-(-1)^q u\right)\,dt,\\[6pt]
d\theta &= \left[\frac{s(u)}{r}\sin\!\left(\theta-(-1)^q u\right)+a\right]dt
+ \sigma\, dW,
\end{aligned}
\right.\label{stoeq}
\end{equation}
where $r$ is the distance to the goal $\tau$, $\theta$ is the counterclockwise angle between the wind direction and LoS to $\tau$, $a$ represents wind drifting, $\sigma$ is the Brownian diffusion intensity, and $W$ is a Wiener process. Speed of vessel is denoted by $u$, $ s(u)$ is  a polar plot relating apparent wind angle and the vessel's theoretical speed, shown in Fig.~\ref{c5_polar}. The polar speed function is not smooth near the upwind region because the boat cannot sail inside the no-go zone unless switch between tacks, producing a piecewise, non-differentiable relationship.

To compute the minimum expected time-to-goal, the sailing domain is discretized into a grid and a Bellman update is applied to obtain a value map over all states and tack configurations. As derived in \cite{sto}, the value function is obtained from a quasi-variational inequality using the Ito--Taylor expansion, yielding the minimum expected arrival time under stochastic wind and tack-switching decisions. Although computationally efficient, this planner is not directly suitable for real-world deployment. The point-mass approximation neglects the inertia and actuator constraints are not explicitly considered, which may lead to aggressive and infeasible trajectories.

\subsection{Baseline Controller}
To improve the feasibility, the reference trajectory generated by \cite{sto} is tracked using NMPC with an explicit sailboat model. NMPC is proposed since it handles nonlinear dynamics and enforces multiple constraints including actuator constraints. The NMPC problem is formulated as Eq.~\eqref{baselinempc}
\begin{equation}
\begin{aligned}
\min_{\mathbf{u},\mathbf{X}} \quad & J=\lVert X(t)-X_d(t)\rVert^2_Q +\lVert \textbf{u}(t)\rVert^2_R\\
\text{s.t.} \quad & \dot{X}(t)=f(X(t),\textbf{u}(t)), \quad t\in[0,T],\\
& \textbf{u}(t)\in U, \quad t\in[0,T],\\
& X(T)\in X_f,
\end{aligned}\label{baselinempc}
\end{equation}
where $X(t)$ is the sailboat state, $X_d(t)$ is the reference state from the baseline planner, $\textbf{u}(t)$ denotes the control inputs (rudder and sail angles, $\delta_r,\delta_s$), $U$ is the admissible input set, and $X_f$ is the terminal constraint set. $Q$ and $R$ are positive definite weighting matrices related to states and control inputs term, respectively. This formulation reduces the mismatch between the point-mass planner and the physical vessel model while penalizing excessive control effort, enabling reliable trajectory realization under stochastic wind disturbances.

\section{Controller Design}
\subsection{Controllability Analysis}
As indicated by the polar plot in Fig.~\ref{c5_polar}, sailboat maneuverability may degrade significantly in certain upwind operating regions, resulting in low speed and limited heading authority. Under stochastic wind disturbances, these conditions can further compromise the vessel's ability to execute reliable turns and tack transitions. Therefore, it is necessary to explicitly detect such maneuvering-degraded regions and prevent the controller from operating there.

To this end, a controllability analysis is developed based on the nonlinear control theory \cite{boscain2019}. Consider a control-affine system:
\begin{equation}
\dot{x} = f_0(x) + \sum_{i=1}^{n} f_i(x)u_i,\label{affine}
\end{equation}
where $f_0(x)$ denotes the drift vector field and $f_i(x)$ are the control vector fields. The Lie algebra rank condition (LARC) requires that the vector fields $f_0,\dots,f_n$ and their iterated Lie brackets span the tangent space at $x$. For systems with drift, LARC guarantees local accessibility, i.e., the set reachable from $x$ has a nonempty interior \cite{sussmann1972,boscain2019}; it does not by itself guarantee controllability, which requires additional properties such as recurrence of the drift \cite{boscain2019}. Since the sailboat drift contains dissipative hydrodynamic damping, such recurrence cannot be assumed, and no controllability certificate is claimed in this paper. Instead, the Lie-algebraic analysis is used as a local measure of control authority. For small times $t$, the motion generated along the input fields is of order $t$, whereas the motion along a bracket of length $k$ is of order $t^{k}$ \cite{boscain2019}. States at which some direction, in particular the surge direction, can no longer be produced by the inputs and their first-order brackets with the drift, but only through higher-order brackets, therefore correspond to a severe reduction of maneuverability even though accessibility may be preserved. Throughout the paper, \emph{controllability-aware} and \emph{low-controllability} are used in this sense.

\subsubsection{Affine reformulation with virtual inputs}
The sailboat maneuvering model is nonlinear and non-affine with respect to the physical actuation variables. To facilitate Lie algebra analysis, the model is written into an affine form by introducing generalized forces $\tau$ as virtual inputs:
\begin{equation}
\left\{
	\begin{aligned}
		\dot\eta &=J(\eta)v,\\
		M\dot v&=-IC(v)v-I[D(\eta,v)+g(\eta)]+I\tau,\\
		\tau &= k_x \tau_{\text{aug},x}+k_y \tau_{\text{aug},y}+k_\phi \tau_{\text{aug},\phi}+k_\psi \tau_{\text{aug},\psi},
	\end{aligned}\label{dynam}
\right.
\end{equation}
where $\eta \text{ and }v$ are the displacements and velocities in the body frame, $M,~ J(\eta),~C(v),~D(\eta, v) \text{ and } g(\eta)$ represent inertial matrix, position-velocity transformation matrix, Coriolis and Centripetal matrix, damping matrix, and restoring torque, respectively. $\tau_{aug,(\cdot)}$ denotes the corresponding actuator contributions, and $k_{(\cdot)}$ is the effectiveness coefficient that scales the available control authority, with $k_{(\cdot)}=0$ indicating loss of actuation in that direction.

\begin{remark}[Virtual and physical inputs]\label{rem:virtual}
The physical inputs are the sail and rudder angles $(\delta_s,\delta_r)$, which enter the dynamics only through the generalized forces $\tau(\textbf{x},\delta_s,\delta_r)$. Consequently, the input directions of the physical system, as well as their first-order brackets with the drift, lie in the span of the virtual-input fields $\textbf{f}_1,\dots,\textbf{f}_4$ and of their brackets $\textbf{f}_5,\dots,\textbf{f}_8$ defined below. Any rank loss detected for the virtual-input system is therefore also a rank loss for the physical system, whereas the converse does not hold. The degradation conditions derived below are thus necessary conditions for the physical system to retain full first-order authority: the states they flag are degraded for the physical system as well, but avoiding them does not by itself certify that the two physical inputs generate every direction.
\end{remark}

Let the state be defined as $\textbf{x}=[x,y,\phi,\psi,u,v,p,r]^\top\in\mathbb{R}^8$, representing surge, sway, roll, yaw and their corresponding rates, and the dynamics expressed in control-affine form:
\begin{equation}
\dot{\textbf{x}}=\textbf{f}_0(\textbf{x})+\textbf{f}_1\tau_{\text{aug},x}+\textbf{f}_2\tau_{\text{aug},y}+\textbf{f}_3\tau_{\text{aug},\phi}+\textbf{f}_4\tau_{\text{aug},\psi},
\end{equation}
where $\textbf{f}_0$ denotes the drift vector field, 
\[
\bm{\textbf{f}_0} =
\begin{bmatrix}
u \cos\psi - v \sin\psi \cos\phi \\

u \sin\psi + v \cos\phi \cos\psi \\

p \\

r \cos\phi \\

-\dot{Y}_v rv + mrv + \sin\alpha_k - \cos\alpha_h - \cos\alpha_k \\

\dot{X}_u ru - mru + \sin\alpha_h \cos\phi + \sin\alpha_k - \cos\alpha_k \\

-\phi^2 - \phi - \sin\alpha_h \cos\phi + \sin\alpha_k - \cos\alpha_k \\

-\dot{X}_u uv + \dot{Y}_v uv
-\sin(\alpha_k-\sin\alpha_h\cos\phi) - \cos\alpha_k
\end{bmatrix}
\]
and $\textbf{f}_1=[0,0,0,0,k_x,0,0,0]^\top$, 
$\textbf{f}_2=[0,0,0,0,0,k_y,0,0]^\top$,
$\textbf{f}_3=[0,0,0,0,0,0,k_\phi,0]^\top$, and
$\textbf{f}_4=[0,0,0,0,0,0,0,k_\psi]^\top$ are constant input vector fields.

\subsubsection{Lie bracket computation}
The Lie bracket between two vector fields $f(x)$ and $g(x)$ is defined as
\begin{equation}
[f,g]_L(x)=\nabla g(x)\,f(x)-\nabla f(x)\,g(x),
\end{equation}
where $\nabla f(x)$ denotes the Jacobian matrix of $f(x)$. Since $\textbf{f}_1$, $\textbf{f}_2$, $\textbf{f}_3$, and $\textbf{f}_4$ are constant vectors, their Jacobians satisfy $\nabla \textbf{f}_1=\nabla \textbf{f}_2=\nabla \textbf{f}_3=\nabla \textbf{f}_4 = 0_{8\times 8}.$ Therefore, the first-order Lie brackets with the drift term reduce to
\begin{equation}
[\textbf{f}_0,\textbf{f}_i]_L(x) = -\nabla \textbf{f}_0(x)\,\textbf{f}_i,\quad i\in\{1,2,3,4\}.
\end{equation}
This provides a computationally efficient procedure: each bracket $[\textbf{f}_0,\textbf{f}_i]_L$ is obtained by multiplying the Jacobian of the drift $\nabla \textbf{f}_0$ with the corresponding constant direction $\textbf{f}_i$.

The corresponding partial derivatives from the full drift term $f_0$, the first-order Lie brackets, $\textbf{f}_5 \triangleq [\textbf{f}_0,\textbf{f}_1]_L,\quad \textbf{f}_6 \triangleq [\textbf{f}_0,\textbf{f}_2]_L, \quad \textbf{f}_7 \triangleq [\textbf{f}_0,\textbf{f}_3]_L, \quad \textbf{f}_8 \triangleq [\textbf{f}_0,\textbf{f}_4]_L $, are computed as:
\[
\begin{array}{@{}c@{\qquad}c@{}}
\textbf{f}_5=
\begin{bmatrix}
-k_x \cos\psi\\
-k_x \sin\psi\\
0\\
0\\
0\\
k_x(m-\dot X_u)r\\
0\\
k_x(\dot X_u-\dot Y_v)v
\end{bmatrix},
&
\textbf{f}_6=
\begin{bmatrix}
k_y \sin\psi\cos\phi\\
-k_y \cos\phi\cos\psi\\
0\\
0\\
k_y(\dot Y_v-m)r\\
0\\
0\\
k_y(\dot X_u-\dot Y_v)u
\end{bmatrix},
\\[6pt]
\textbf{f}_7=
\begin{bmatrix}
0\\0\\-k_\phi\\0\\0\\0\\0\\0
\end{bmatrix},
&
\textbf{f}_8=
\begin{bmatrix}
0\\0\\0\\-k_\psi\cos\phi\\
k_\psi(\dot Y_v v-mv)\\
k_\psi(-\dot X_u u+mu)\\
0\\0
\end{bmatrix}.
\end{array}
\]

In addition, Lie brackets among the constant input vector fields are identically zero:
\[
[\textbf{f}_i,\textbf{f}_j]_L(x)=0_{8\times 1},\quad i,j\in\{1,2,3,4\},
\]
so that brackets among input fields alone generate no new direction. Assembling the input fields and their first-order brackets with the drift into $M_1(\textbf{x})=[\textbf{f}_1,\dots,\textbf{f}_8]\in\mathbb{R}^{8\times 8}$, a symbolic computation gives
\[
\det M_1(\textbf{x}) = k_x^2\,k_y^2\,k_\phi^2\,k_\psi^2\cos^2\phi .
\]
Since $|\phi|<\pi/2$ during sailing, $M_1$ has full rank, and LARC is satisfied with first-order brackets only, if and only if all effectiveness coefficients are nonzero. When one of them vanishes, the first-order rank is lost and higher-order brackets must be examined. For instance, at the surge-degenerate configuration characterized below ($k_x=0$, $v=r=0$), the input fields and first-order brackets span only a six-dimensional subspace, while adding the second-order brackets restores rank eight (verified symbolically). The system therefore remains locally accessible, but the lost directions can only be generated through second-order brackets, i.e., through slow and indirect coupling effects. Consequently, the criterion adopted in this paper is the loss of first-order rank, rather than the failure of LARC.

\subsubsection{Controllability degradation conditions}
To interpret the rank loss physically, consider $M_0(\textbf{x})=[\textbf{f}_0,\textbf{f}_1,\dots,\textbf{f}_8]\in\mathbb{R}^{8\times 9}$. Its fifth row collects every component along the surge acceleration $\dot u$ that can be produced by the drift, by the inputs, or by their first-order brackets. If this row vanishes, $\mathrm{rank}\,M_0<8$ and the surge acceleration cannot be influenced at first order by any mechanism, neither directly by the sail nor indirectly through the sway--yaw coupling. This vanishing of the surge-propulsion channel is the most critical degradation for a sailboat, since forward speed is required for rudder effectiveness and hence for all other maneuvers. From the expressions of $\textbf{f}_0,\dots,\textbf{f}_8$, the fifth row of $M_0$ vanishes if and only if
\begin{align}
\hspace{-1.2em}k_x &=0, \label{tau}\\
\hspace{-1.2em}-\dot{Y}_v rv + mrv + \sin(\alpha_{ak}) - \cos(\alpha_{ah}) - \cos(\alpha_{ak}) &=0, \label{p}\\
\hspace{-1.2em}k_y(\dot{Y}_v r - mr) &=0, \label{r}\\
\hspace{-1.2em}k_\psi(\dot{Y}_v v - mv) &=0. \label{v}
\end{align}

As long as the rudder retains authority in sway and yaw ($k_y\neq0$, $k_\psi\neq0$), and since $\dot Y_v\neq m$ for any physical hull, conditions \eqref{r} and \eqref{v} hold if and only if $r=0$ and $v=0$. Hence, $v=r=0$ is not a simplifying assumption: it is implied by \eqref{r}--\eqref{v}, and substituting it into \eqref{p} neither relaxes nor tightens the degeneracy condition. The surge-degenerate set is therefore exactly characterized by $k_x=0$, $v=r=0$, and \eqref{p} evaluated at $v=r=0$. Expressing the apparent angles in terms of the state, the latter condition delimits the maneuvering-degraded region
\begin{equation}
p \ge u(\cos\phi + 1).\label{degraded}
\end{equation}

The effectiveness coefficient $k_x$ quantifies the net surge authority provided by the sail and the rudder. It is modeled as
\begin{equation}
k_x = \kappa\big(\tau_x - F_{rh}\big),\quad \kappa(s)=0 \text{ if } s\le 0,\quad \kappa(s)>0 \text{ if } s>0,\label{kx_def}
\end{equation}
where $\tau_x=\tau_{s,x}+\tau_{r,x}$ is the surge generalized force produced by the sail and the rudder, and $F_{rh}$ is the hull resistance. In words, the surge channel is considered actuated only when the propulsive force exceeds the resistance; otherwise, the actuators can only slow the vessel down in surge. This relation is a physically motivated modeling assumption, not a consequence of the Lie-algebraic analysis; its role is to express the abstract condition \eqref{tau} in terms of forces that can be evaluated online, as detailed below.

\paragraph{Thrust decomposition}
The net surge generalized force $\tau_x$ is decomposed into the aerodynamic contribution from the sail and the hydrodynamic contribution from the rudder, $\tau_x=\tau_{s,x}+\tau_{r,x}.$

The sail generates both lift $L_s$ and drag $D_s$ in the apparent-wind frame, which can be projected onto the body-frame surge axis using the apparent wind angle of attack $\alpha_{aw}$:
\begin{equation}
\tau_{s,x}=L_s\sin\alpha_{aw}-D_s\cos\alpha_{aw}.
\end{equation}

The rudder contribution is dominated by drag in the surge direction and is modeled as $\tau_{r,x}=-D_r$. Therefore, the total surge force can be written as
\begin{equation}
\tau_x=L_s\sin\alpha_{aw}-D_s\cos\alpha_{aw}-D_r.
\label{eq:tau_x_basic}
\end{equation}

\paragraph{Aerodynamic and hydrodynamic force models}
The sail lift and drag, and the rudder drag, are modeled using standard quadratic forms:
\begin{equation}
\begin{aligned}
L_s &= \frac{1}{2}\rho_a A_s V_{aw}^2 C_{L_s}(\alpha_{aw}),\\
D_s &= \frac{1}{2}\rho_a A_s V_{aw}^2 C_{D_s}(\alpha_{aw}),\\
D_r &= \frac{1}{2}\rho_w A_r V_{ar}^2 C_{D_r}(\alpha_{ar}),
\end{aligned}
\label{eq:aero_hydro_models}
\end{equation}
where $\rho_a$ and $\rho_w$ denote the densities of air and water, respectively. $A_s$ and $A_r$ are the effective sail and rudder reference areas. $V_{aw}$ is the apparent wind speed experienced by the sail, and $V_{ar}$ is the apparent inflow speed at the rudder. The coefficients $C_{L_s}(\cdot)$, $C_{D_s}(\cdot)$, and $C_{D_r}(\cdot)$ are obtained by curve fitting empirical or identified force data.

\paragraph{Explicit surge feasibility inequality}
Substituting the models of Eq.~\eqref{eq:aero_hydro_models} into Eq.~\eqref{eq:tau_x_basic} yields:
\begin{equation*}
\begin{aligned}
\tau_x
&=
\frac{1}{2}\rho_a A_s V_{aw}^2
\Big(
C_{L_s}(\alpha_{aw})\sin\alpha_{aw}
-
C_{D_s}(\alpha_{aw})\cos\alpha_{aw}
\Big)\\
&-
\frac{1}{2}\rho_w A_r V_{ar}^2
C_{D_r}(\alpha_{ar}).
\end{aligned}
\label{eq:tau_x_explicit}
\end{equation*}

The hull resistance $F_{rh}$ is approximated as a polynomial function of the apparent hull-relative speed $v_{ah}$:
\begin{equation*}
F_{rh}=a_h v_{ah}^3+b_h v_{ah}^2+c_h v_{ah}+d_h,
\label{eq:Frh}
\end{equation*}
where $a_h,b_h,c_h,d_h$ are identified coefficients.

Hence, the loss-of-propulsion condition $k_x=0$ can be interpreted as:
\begin{equation*}
\begin{aligned}
&\frac{1}{2}\rho_a A_s V_{aw}^2
\Big(
C_{L_s}(\alpha_{aw})\sin\alpha_{aw}
-
C_{D_s}(\alpha_{aw})\cos\alpha_{aw}
\Big)\\
&-
\frac{1}{2}\rho_w A_r V_{ar}^2
C_{D_r}(\alpha_{ar})
\le
a_h v_{ah}^3+b_h v_{ah}^2+c_h v_{ah}+d_h,
\label{eq:surge_feasibility}
\end{aligned}
\end{equation*}
which provides an explicit physical interpretation: the sailboat can only achieve net forward motion when the surge direction aerodynamic thrust generated by the sail exceeds the combined hydrodynamic drag from the rudder and the resistive forces from the hull.

\paragraph{Implementation in NMPC}
By \eqref{kx_def}, the strict inequality
\begin{equation}
\tau_x(t) > F_{rh}(t), \quad t\in[0,T],\label{surgecon}
\end{equation}
is equivalent to $k_x>0$. It keeps $\textbf{f}_1$ and $\textbf{f}_5$ nonzero along the predicted trajectory, so that the fifth row of $M_0$ cannot vanish and the surge channel retains direct first-order authority. The logical chain is thus: \eqref{surgecon} $\Rightarrow$ $k_x>0$ $\Rightarrow$ \eqref{tau} is violated $\Rightarrow$ the surge-degenerate set is avoided. Constraint \eqref{surgecon} is a propulsion-feasibility condition, and its link with the rank analysis is established through the modeling assumption \eqref{kx_def}.

\subsection{Proposed NMPC Controller}
Based on the above controllability analysis, the NMPC formulation is augmented by additional controllability constraints to avoid maneuvering-degraded regions. The resulting NMPC problem is:
{\small
\begin{equation}
\hspace{-.5cm}
\begin{aligned}
	\min_{\mathbf{u},\mathbf{X}} \quad & J=\lVert X(t)-X_d(t)\rVert^2_Q+\lVert \textbf{u}(t)\rVert^2_R\\
	\text{s.t.}& \\
	&\quad \dot{X}(t) = f(X(t), \textbf{u}(t)) \quad t \in [0, T] \text{ from Eq. \eqref{dynam}}, \\
	&\quad \textbf{u}(t) \in U \quad t \in [0, T], \\
	&\quad X(T) \in X_f ,\\
	&\quad \lVert v(t)\rVert \geq \epsilon_v,\;\; \lVert r(t)\rVert \geq \epsilon_r \quad t \in [0, T], \\
	&\quad p(t) < u(t)(\cos\phi(t)+1) \quad t \in [0, T], \\
	&\quad \tau_x(t) > F_{rh}(t) \quad t \in [0, T],
\end{aligned}\label{modifiedmpc}
\end{equation}
}
where $\epsilon_v$ and $\epsilon_r$ are small positive margins on the sway velocity and yaw rate. Since the surge-degenerate set requires \eqref{tau}--\eqref{v} to hold simultaneously, each of the three additional constraint types excludes it on its own: the margins on $v$ and $r$ violate \eqref{r}--\eqref{v}, the inequality on $p$ excludes the region \eqref{degraded}, and \eqref{surgecon} violates \eqref{tau}. They are enforced jointly to provide redundancy: under stochastic wind, the realized surge force may differ from its prediction, in which case the sway--yaw coupling terms still provide first-order surge authority. These constraints therefore impose a margin away from maneuvering-degraded conditions, rather than a formal controllability guarantee, and improve robustness under stochastic wind.


\label{sec:Controller Design}
\section{Performance Evaluation}
This section evaluates the proposed stochastic wind planning and NMPC tracking framework with the controllability-aware constraints. Side-wind and tacking maneuvers under both deterministic and stochastic wind profiles are first presented. 
The baseline planner and controller are then compared against the proposed controllability-analyzed counterpart. All algorithms are implemented in \textsc{Julia} and interfaced with \textsc{MATLAB} via \textsc{CasADi}. Simulations are conducted on a 12\,m class sailboat model \cite{xiao}. The sampling time is $\delta=0.2$\,s and the NMPC prediction horizon is $T=10\delta$.

\subsection{Side-Wind Sailing Performance}
This subsection examines the trajectory tracking performance in a side-wind scenario under both deterministic and stochastic apparent wind. For the deterministic case, the apparent wind angle and speed are fixed at $\alpha_{aw}=90^\circ$ and $v_{aw}=4$\,m/s. For the stochastic case, $v_{aw}$ remains 4~m/s while $\alpha_{aw}$ evolves from $180^\circ$ to $240^\circ$ within a 10\,s window with random fluctuations. To account for environmental/model uncertainties, wave disturbances are injected into the vessel motion, which are $10\%$ perturbations in surge and sway and $100\%$ perturbations in roll and yaw.

\subsubsection{Deterministic Case}
\begin{figure*}[htbp]
	\subfloat[Trajectory without disturbances\label{c5_side_deter_traj}]{\includegraphics[width=0.495\linewidth]{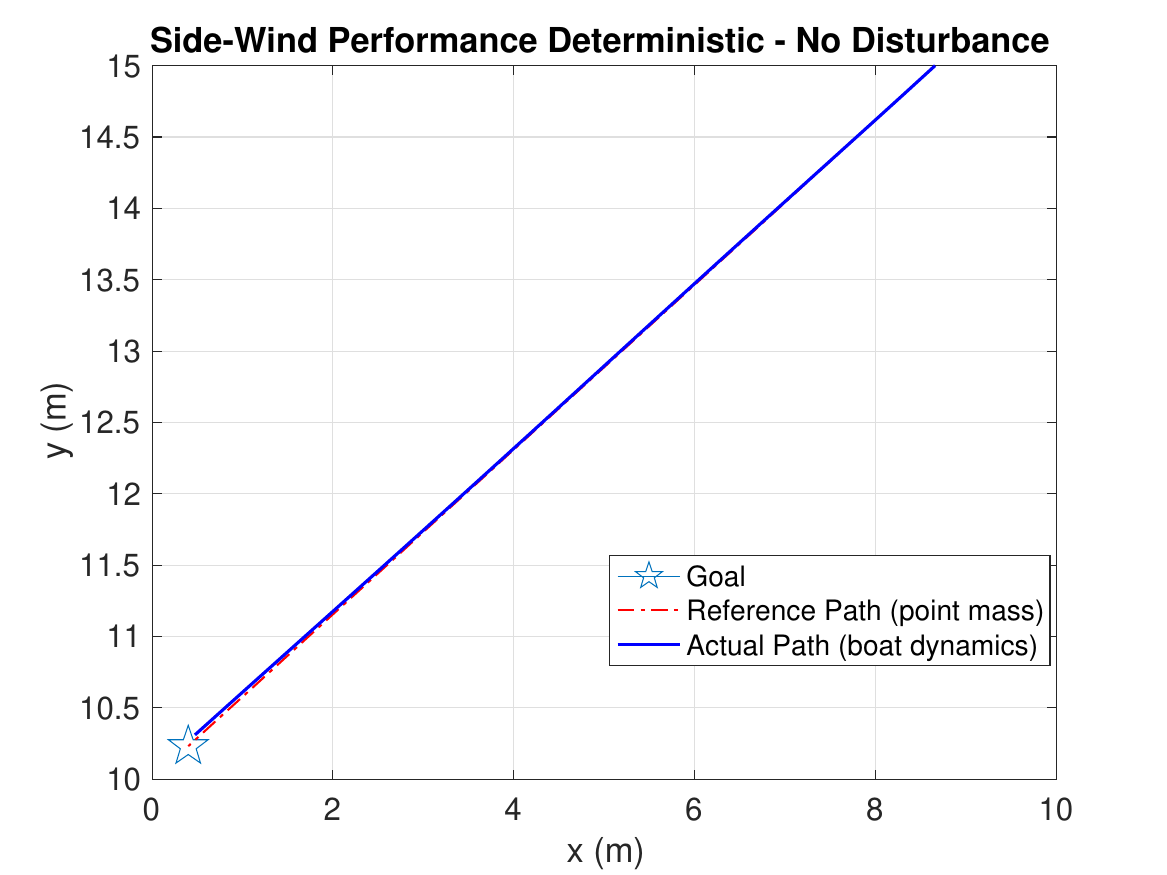}}
	\hfill
	\subfloat[Control inputs without disturbances\label{c5_side_deter_angle}]{\includegraphics[width=0.495\linewidth]{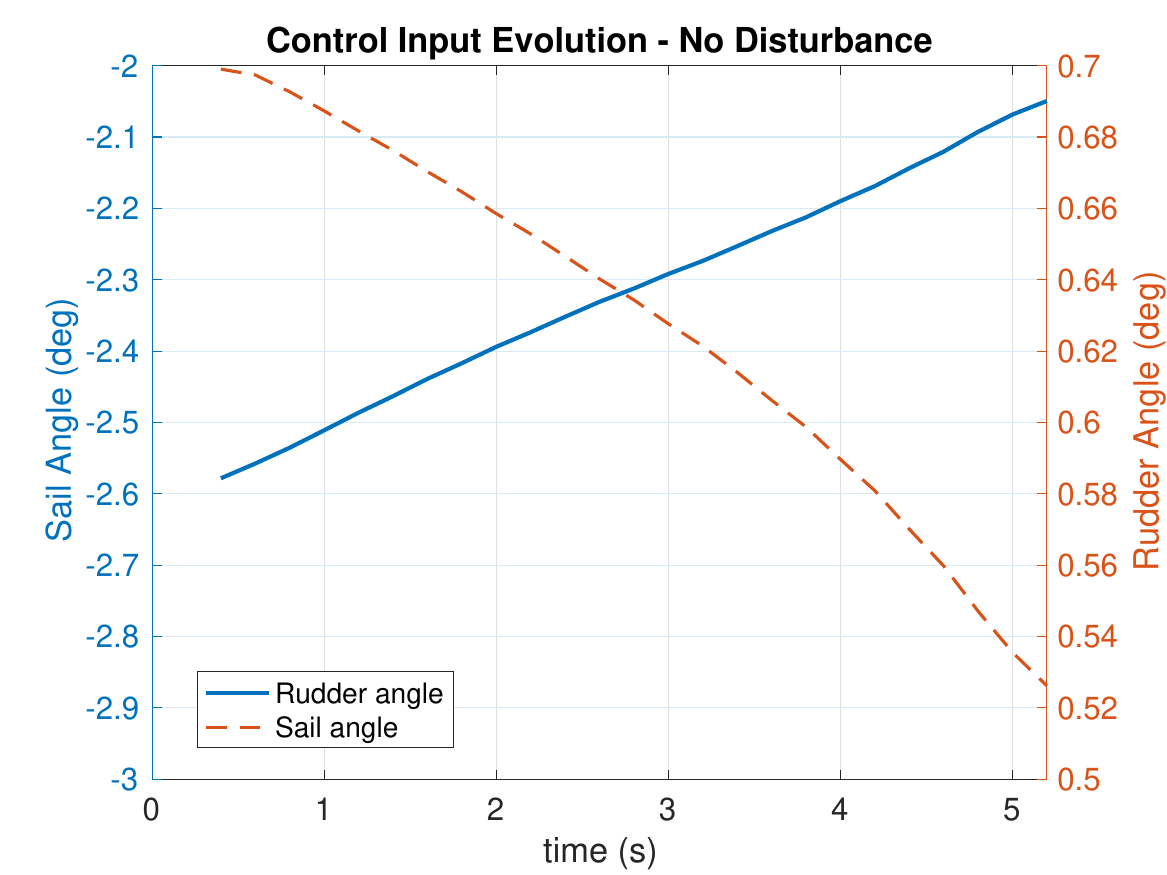}}

	\subfloat[Trajectory with disturbances\label{c5_side_deter_traj2}]{\includegraphics[width=0.495\linewidth]{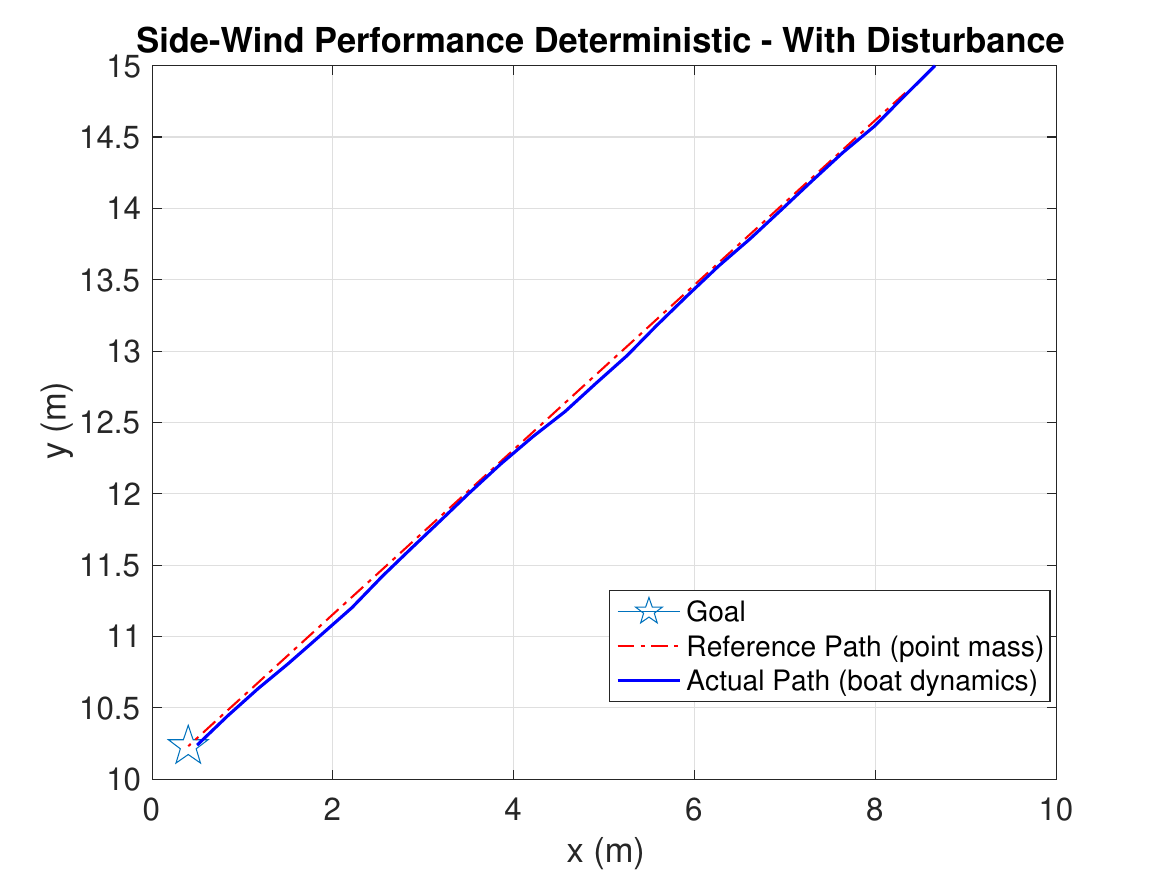}}
	\hfill
	\subfloat[Control inputs with disturbances\label{c5_side_deter_angle2}]{\includegraphics[width=0.495\linewidth]{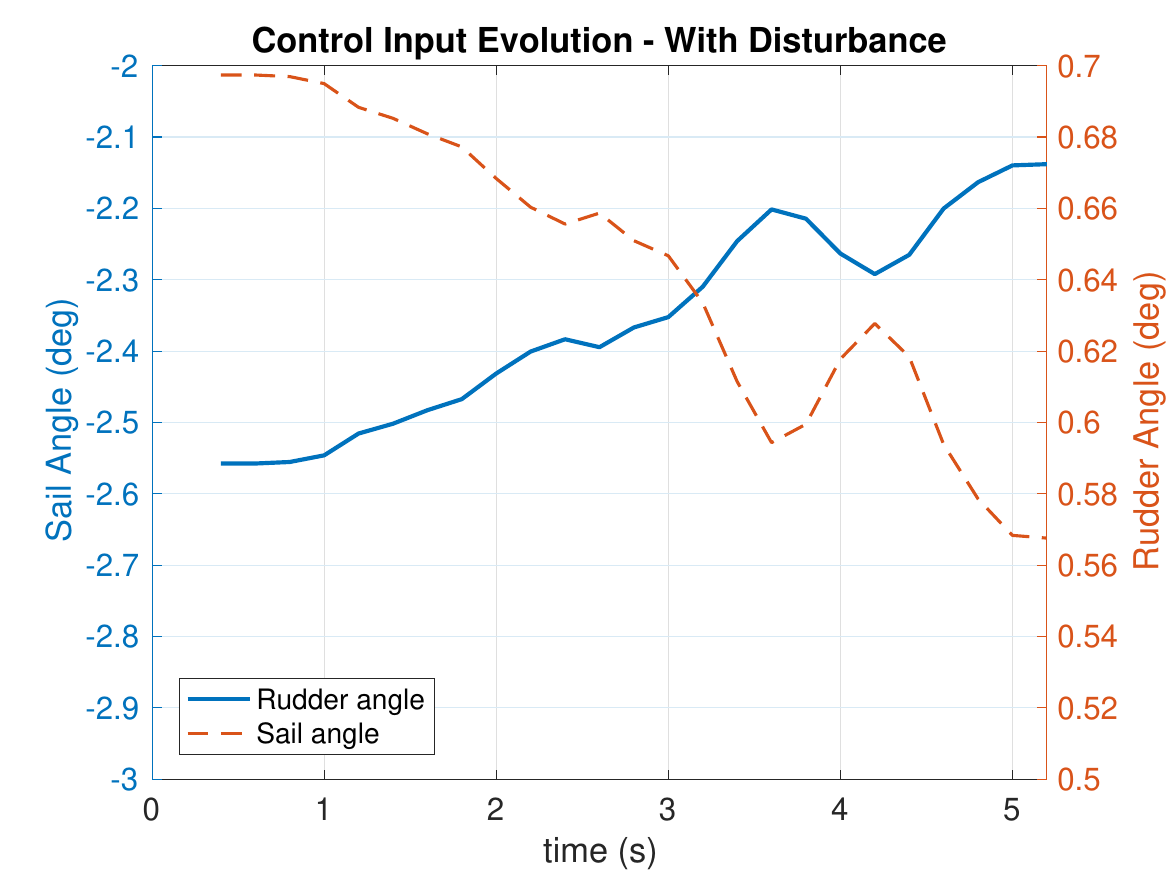}}
	\caption{Side-wind sailing results (deterministic wind).}
	\label{c5_side_deter}
\end{figure*}

Fig.~\ref{c5_side_deter}(a) and Fig.~\ref{c5_side_deter}(b) show that, under deterministic wind (known wind direction and speed) and without external perturbations, the NMPC controller tracks the reference trajectory with high accuracy while keeping rudder and sail control inputs within admissible bounds. The tracking error between the realized and reference paths is small, with a trajectory root mean square error (RMSE) of 0.113\,m. Under injected $10\%$ sway and roll disturbances, as illustrated in Fig.~\ref{c5_side_deter}(c) and Fig.~\ref{c5_side_deter}(d), the controller maintains comparable performance (RMSE of 0.172\,m) by introducing modest input corrections. In all cases, the inputs take effect after the first control interval, which is consistent with the receding-horizon structure of MPC.

\subsubsection{Stochastic Case}
\begin{figure*}[htbp]
	\subfloat[Trajectory without disturbances\label{c5_side_sto_traj}]{\includegraphics[width=0.495\linewidth]{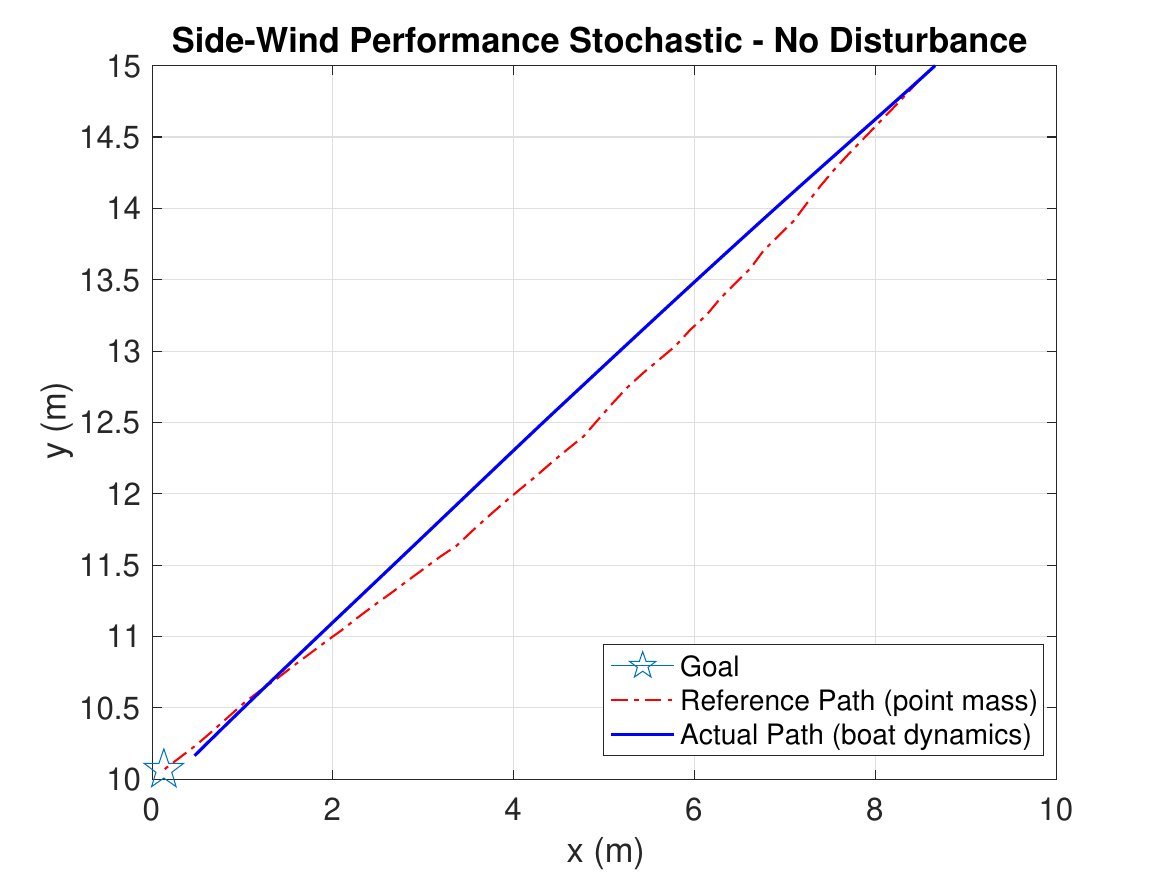}}
	\hfill
	\subfloat[Control inputs without disturbances\label{c5_side_sto_angle}]{\includegraphics[width=0.495\linewidth]{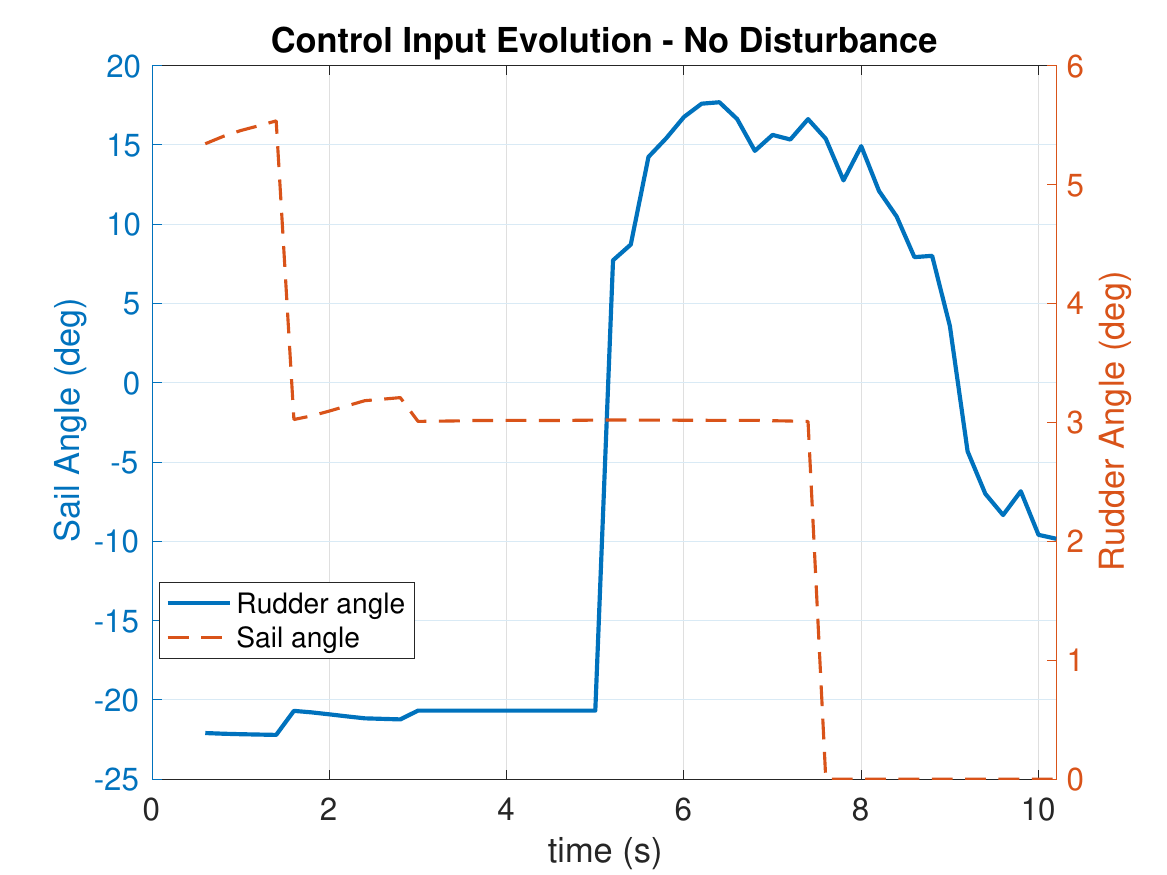}}

	\subfloat[Trajectory with disturbances\label{c5_side_sto_traj2}]{\includegraphics[width=0.495\linewidth]{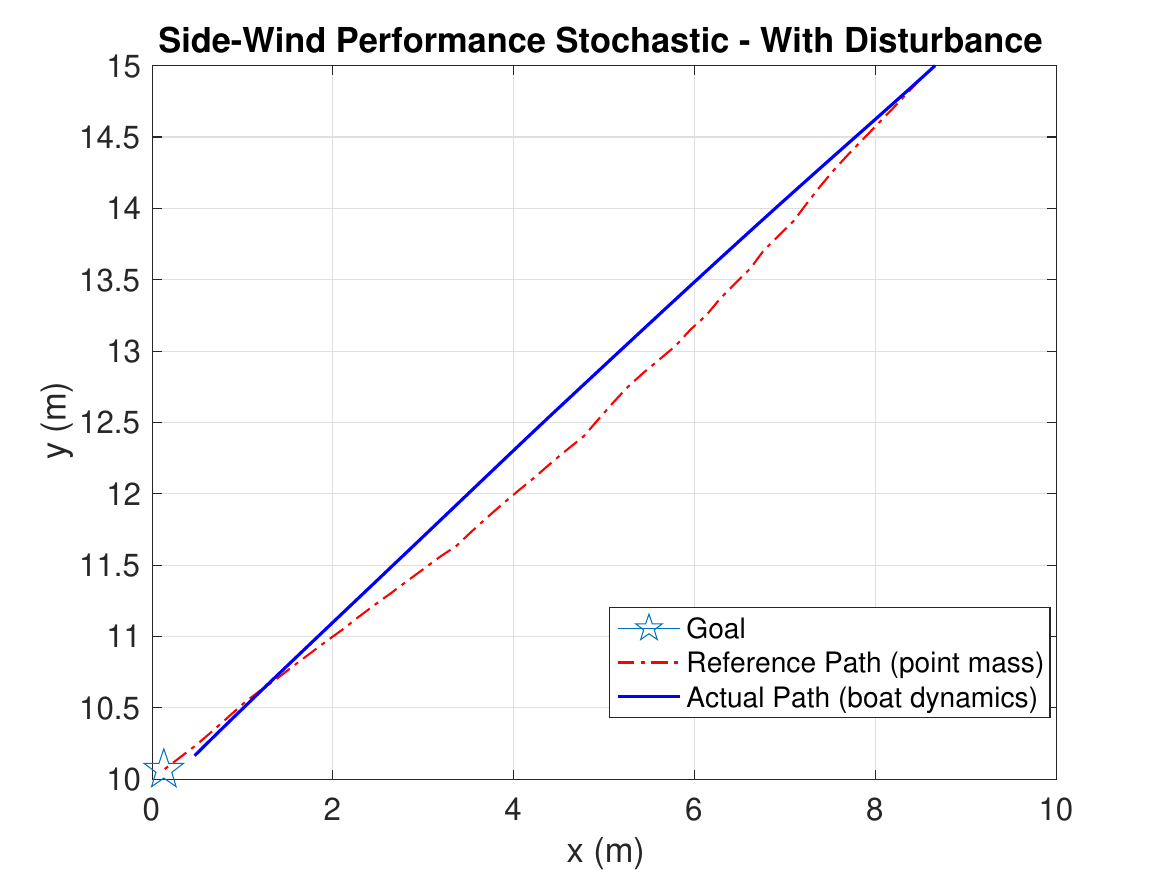}}
	\hfill
	\subfloat[Control inputs with disturbances\label{c5_side_sto_angle2}]{\includegraphics[width=0.495\linewidth]{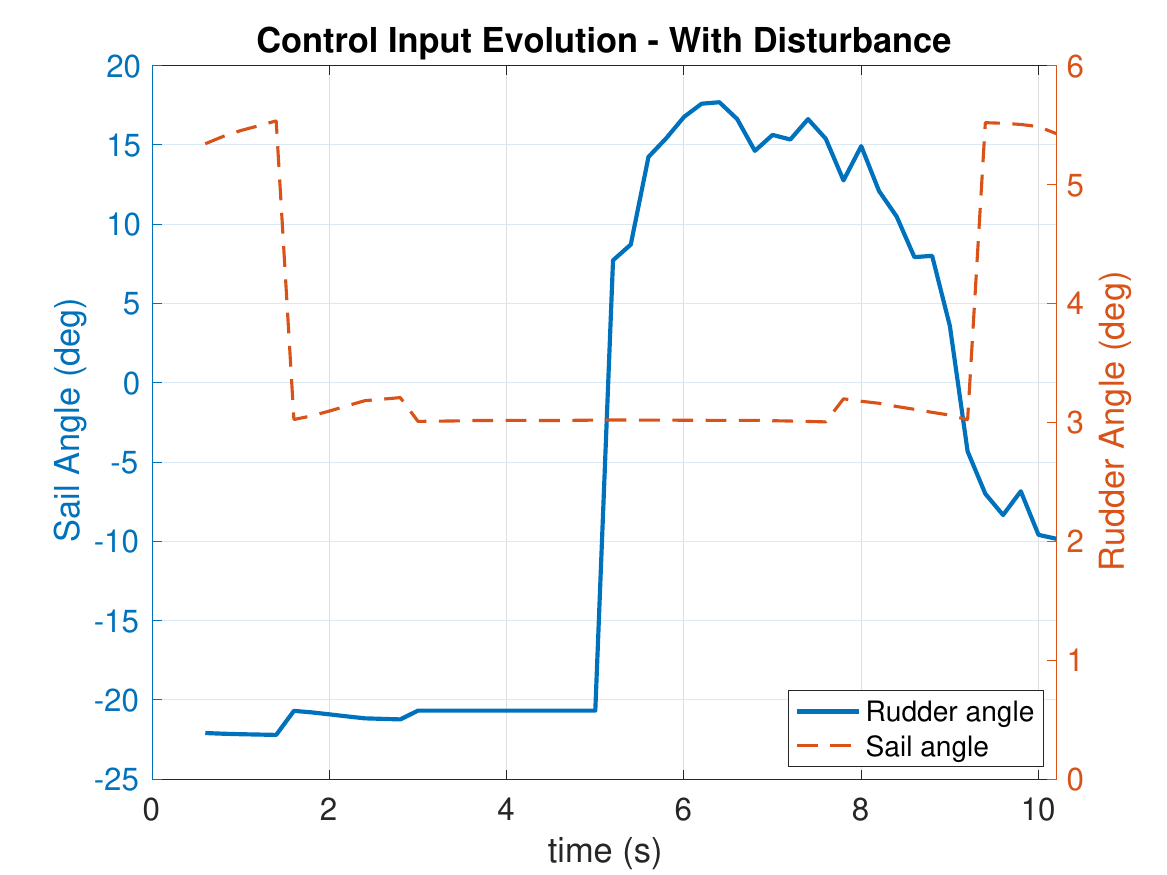}}
	\caption{Side-wind sailing results (stochastic wind).}
	\label{c5_side_sto}
\end{figure*}

Fig.~\ref{c5_side_sto} shows side-wind results under stochastic apparent wind, as formulated in Eq. \eqref{stoeq}. As shown in Figs.~\ref{c5_side_sto}(a) and Fig.~\ref{c5_side_sto}(b), the realized trajectory without wave disturbances remains close to the reference path with RMSE 0.188~m. When wave disturbances are applied, as shown in Fig.~\ref{c5_side_sto}(c) and Fig.~\ref{c5_side_sto}(d), the controller preserves tracking quality (RMSE 0.271~m) by increasing the magnitude and frequency of rudder and sail adjustments. Compared to the deterministic case (Fig.~\ref{c5_side_deter}), the stochastic-wind inputs are noticeably larger, reflecting the NMPC controller's effectiveness in compensating for time-varying wind uncertainties. 


\subsection{Tacking Maneuver Performance}
We next evaluate the proposed controller during tacking maneuvers. The deterministic scenario uses $\alpha_{aw}=0^\circ$ and $v_{aw}=4$\,m/s. In the stochastic scenario, $v_{aw}$ remains 4\,m/s while $\alpha_{aw}$ varies randomly from $180^\circ$ to $285^\circ$ over 27\,s. The disturbance injection follows the same setting used in the side-wind evaluation.

\subsubsection{Deterministic Case}
\begin{figure*}[htbp]
	\subfloat[Trajectory without disturbances\label{c5_tack_deter}]{\includegraphics[width=0.495\linewidth]{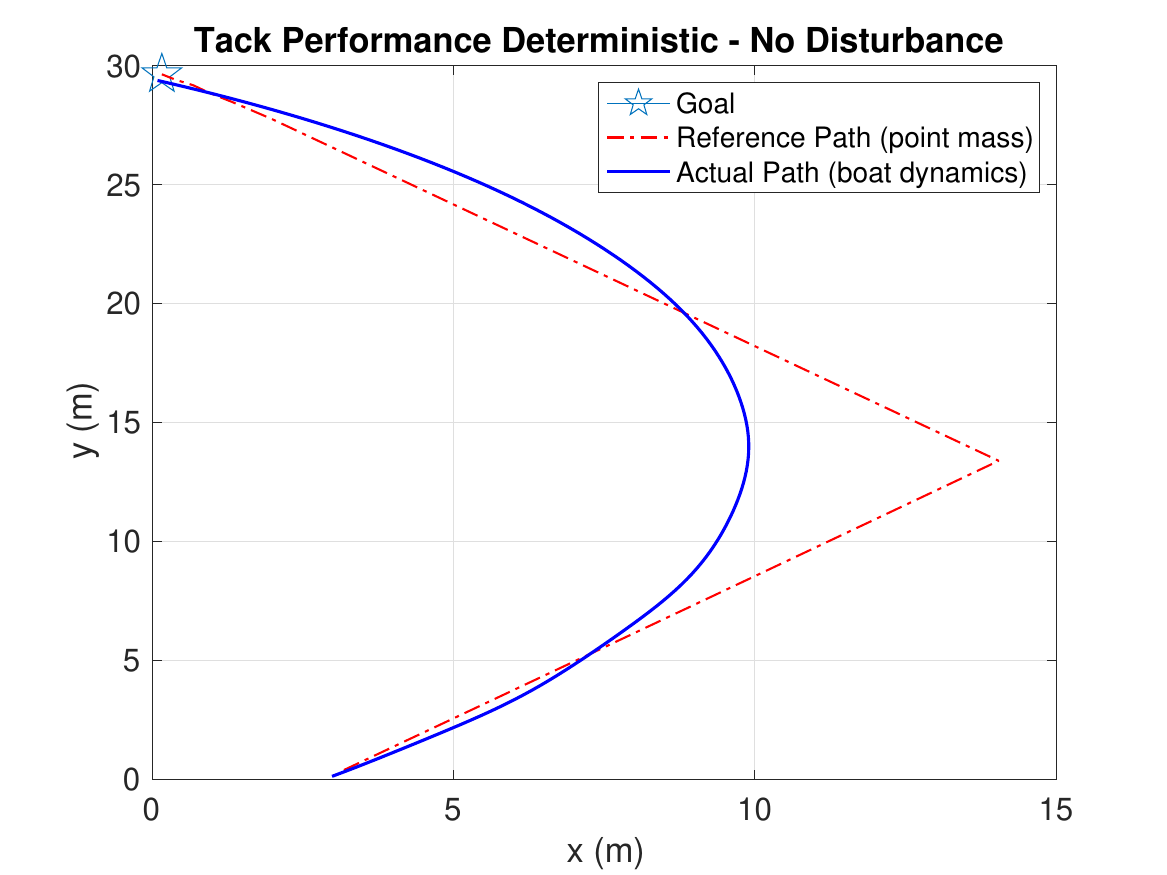}}
	\hfill
	\subfloat[Control inputs without disturbances\label{c5_tack_angle}]{\includegraphics[width=0.495\linewidth]{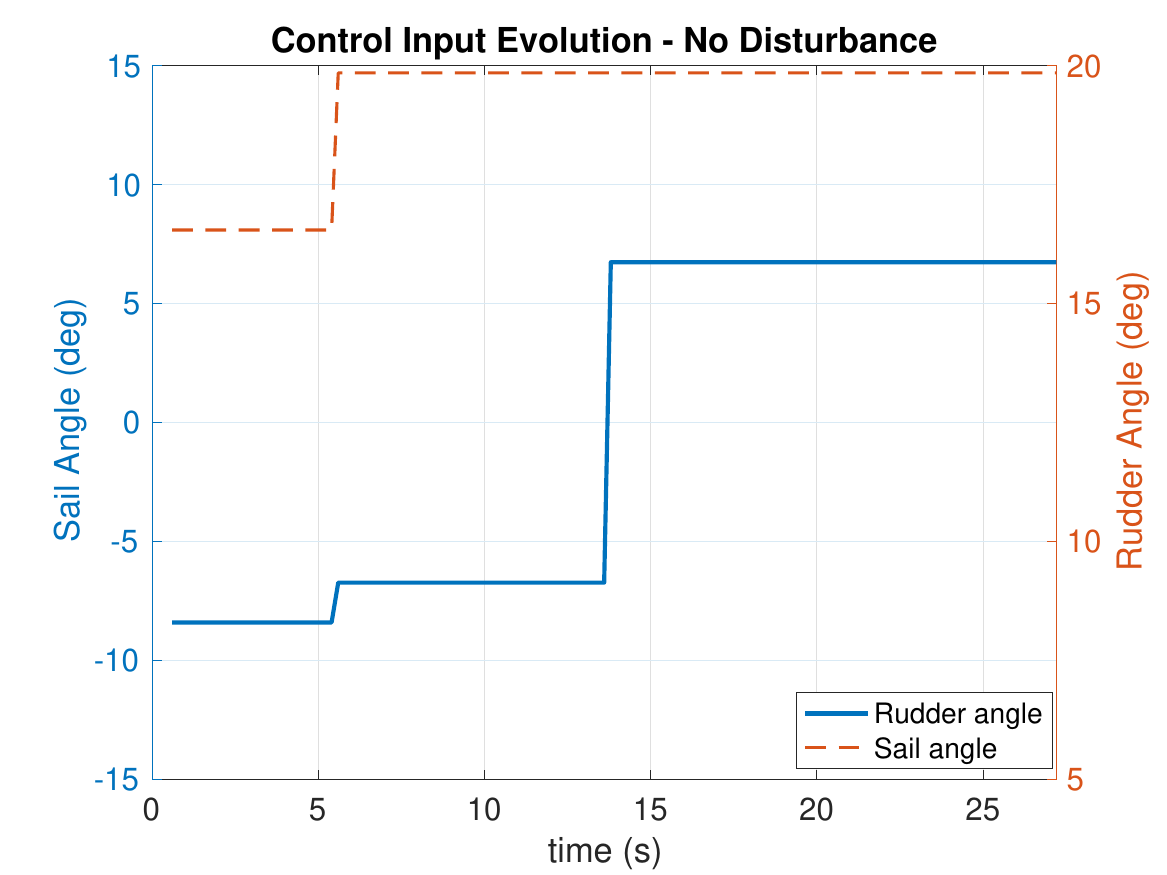}}

	\subfloat[Trajectory with disturbances\label{c5_tack_deter2}]{\includegraphics[width=0.495\linewidth]{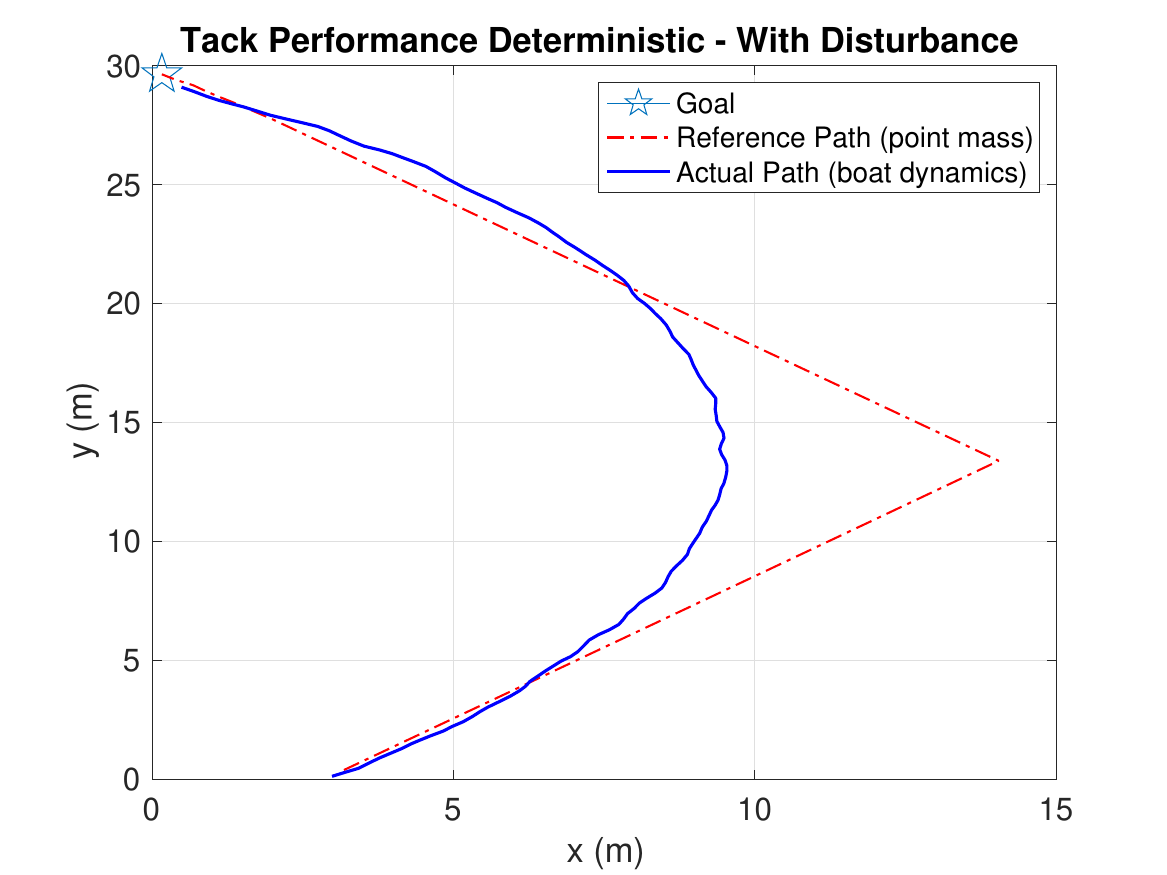}}
	\hfill
	\subfloat[Control inputs with disturbances\label{c5_tack_angle2}]{\includegraphics[width=0.495\linewidth]{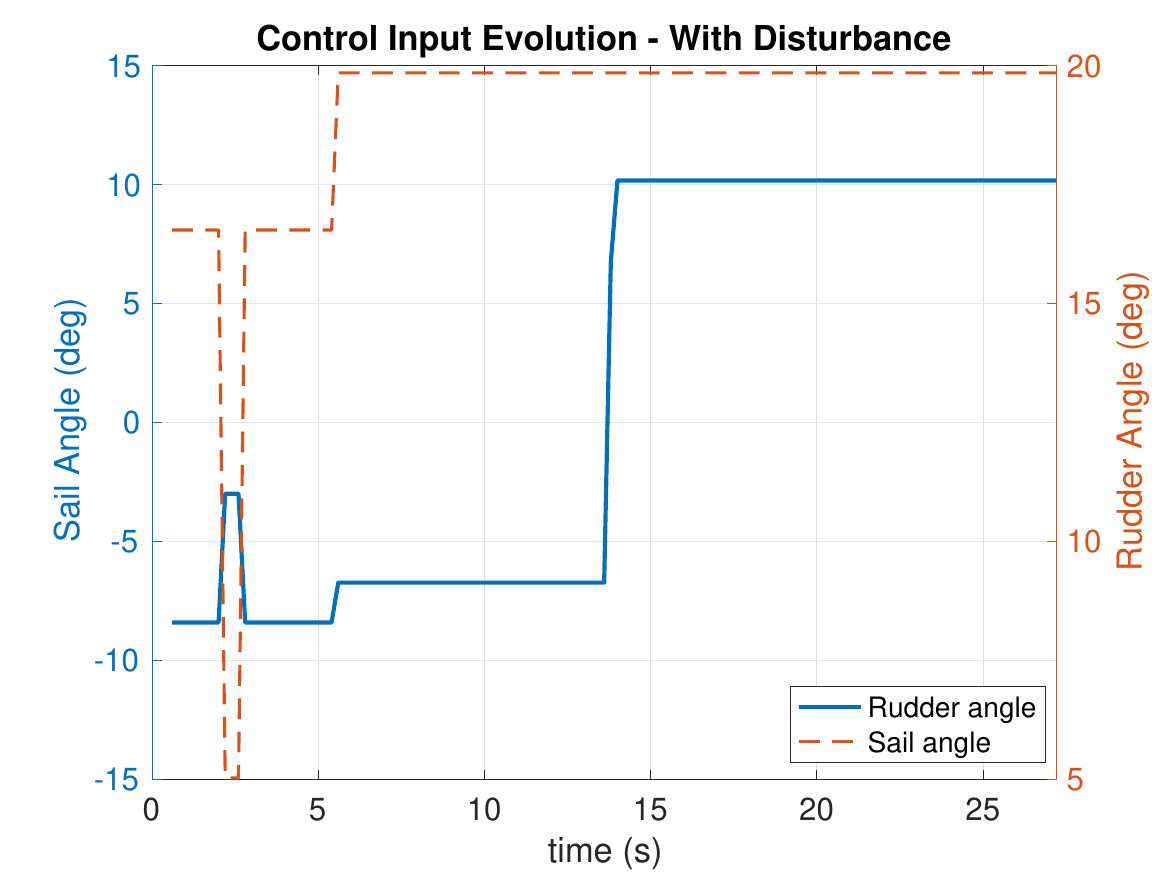}}
	\caption{Tack sailing results (deterministic wind).}
	\label{c5_tack}
\end{figure*}

Fig.~\ref{c5_tack} summarizes deterministic-wind tacking results. Without disturbances, as shown in Fig.~\ref{c5_tack}(a) and Fig.~\ref{c5_tack}(b), the realized trajectory follows the reference path with RMSE 1.87\,m. With disturbances, as shown in Fig.~\ref{c5_tack}(c) and Fig.~\ref{c5_tack}(d), the controller compensates for perturbations through larger initial corrections in rudder and sail commands, while maintaining bounded inputs and stable tracking, with RMSE = 4.96~m.

\subsubsection{Stochastic Case}
\begin{figure*}[htbp]
	\subfloat[Trajectory without disturbances\label{c5_tack_traj_sto}]{\includegraphics[width=0.495\linewidth]{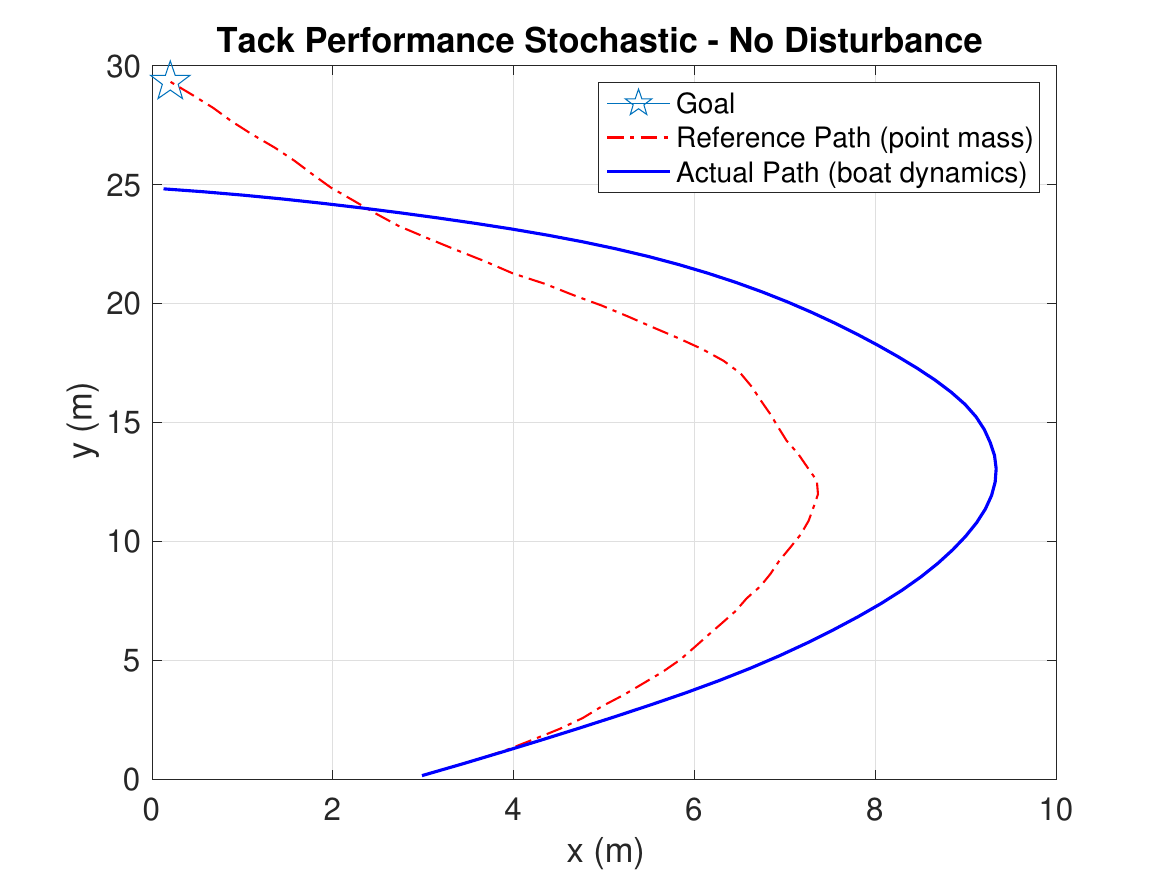}}
	\hfill
	\subfloat[Control inputs without disturbances\label{c5_tack_sto_angle}]{\includegraphics[width=0.495\linewidth]{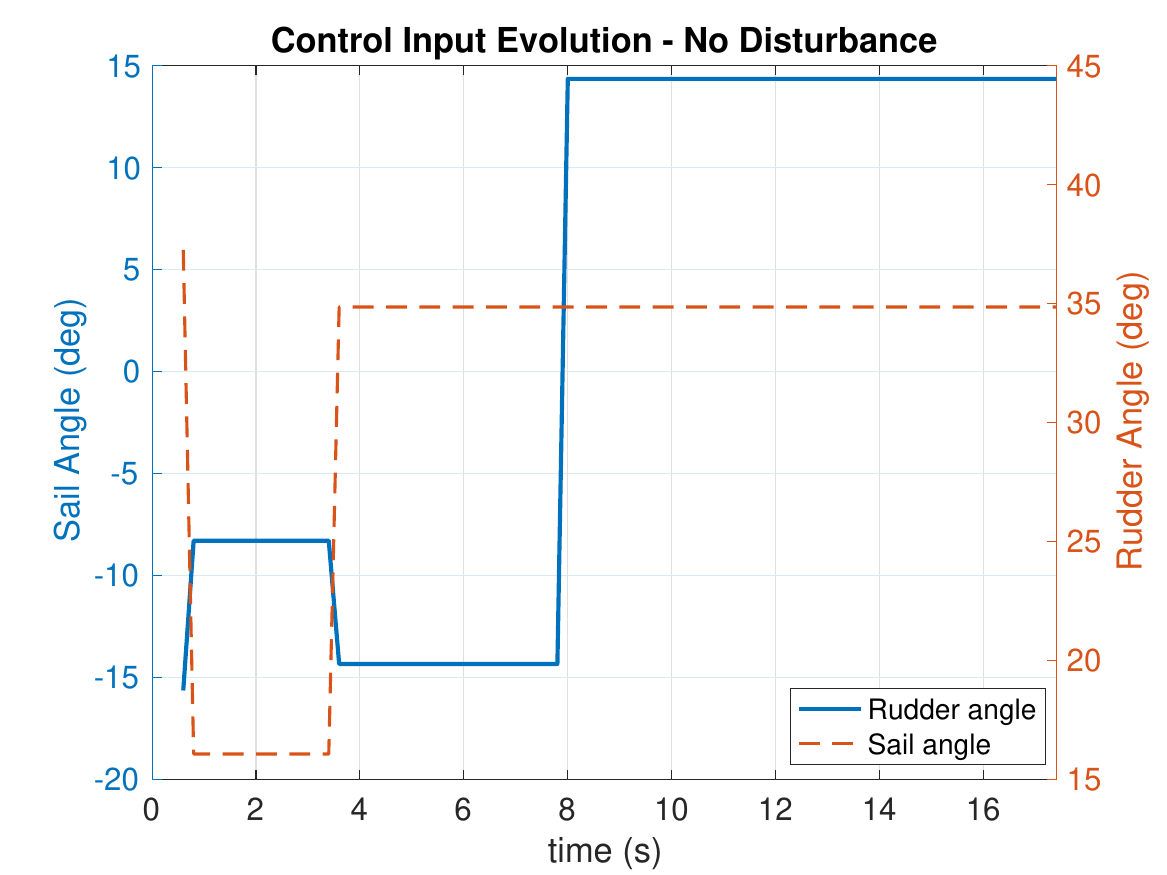}}

	\subfloat[Trajectory with disturbances\label{c5_tack_traj_sto2}]{\includegraphics[width=0.495\linewidth]{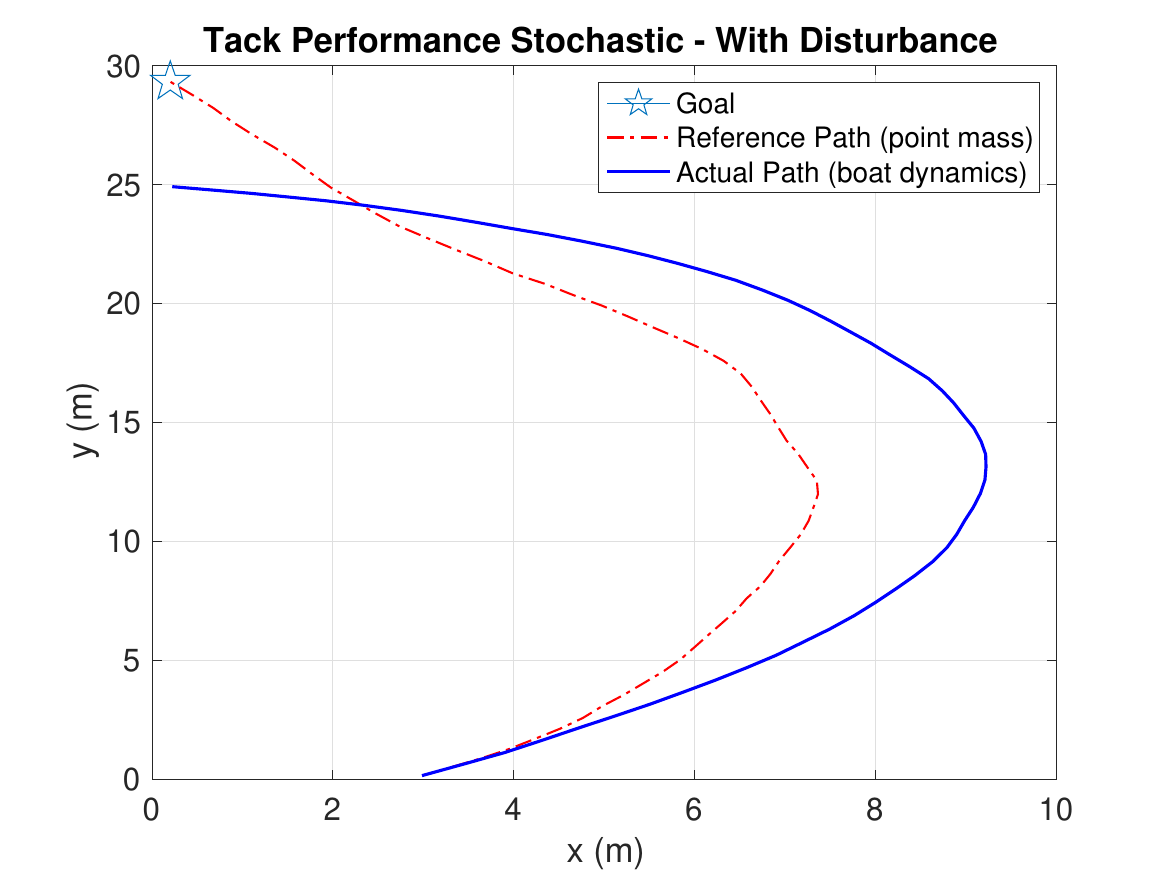}}
	\hfill
	\subfloat[Control inputs with disturbances\label{c5_tack_sto_angle2}]{\includegraphics[width=0.495\linewidth]{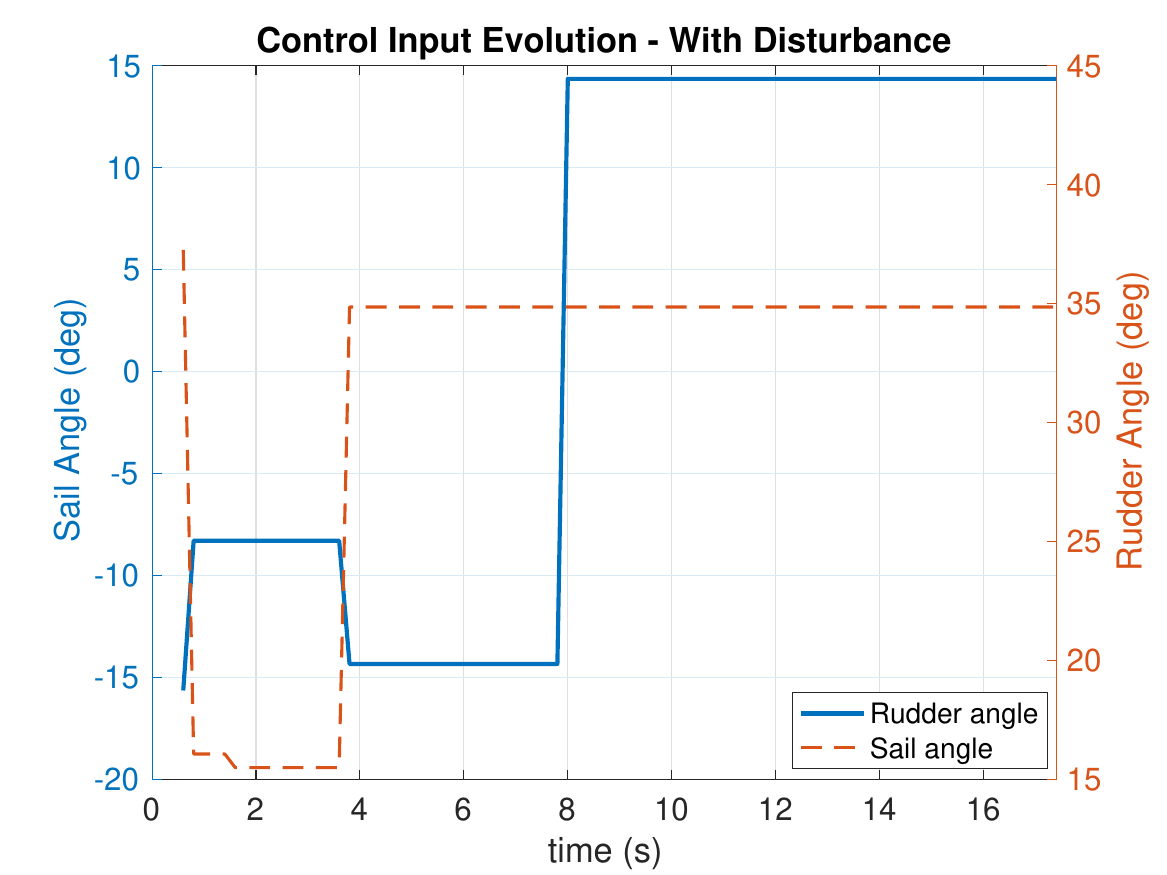}}
	\caption{Tack sailing results (stochastic wind).}
	\label{c5_tack_sto}
\end{figure*}

Under the stochastic wind, shown in Fig.~\ref{c5_tack_sto}, the controller maintains stable tacking performance. Without disturbances, as shown in Fig.~\ref{c5_tack_sto}(a) and Fig.~\ref{c5_tack_sto}(b), the trajectory follows the reference contour with RMSE 2.59~m. When disturbances are present, as illustrated in Fig.~\ref{c5_tack_sto}(c) and Fig.~\ref{c5_tack_sto}(d), the realized path remains similar to the nominal stochastic case and achieves RMSE = 2.61~m. In this simulation, the control inputs remain nearly unchanged, except for a small ($\approx 1^\circ$) rudder reduction between 1.8~s and 3.8~s, suggesting that the NMPC solution stayed within a locally robust region of the reference trajectory. 



\subsection{Comparison of the Proposed Approach with the Baseline Approach}
The baseline planner and controller (``constraints off'') is compared with the proposed controllability-analyzed framework (``constraints on'') in a trajectory tracking subject to strong drift and highly stochastic wind conditions as follows. 
The apparent wind angle $\alpha_{aw}$ varies from $180^\circ$ to $-127^\circ$ over 45~s with stochastic variance $0.2^\circ$.
\begin{figure*}[t]
	\centering
	\subfloat[Trajectory comparison\label{c5_cons_traj}]{\includegraphics[width=0.49\linewidth]{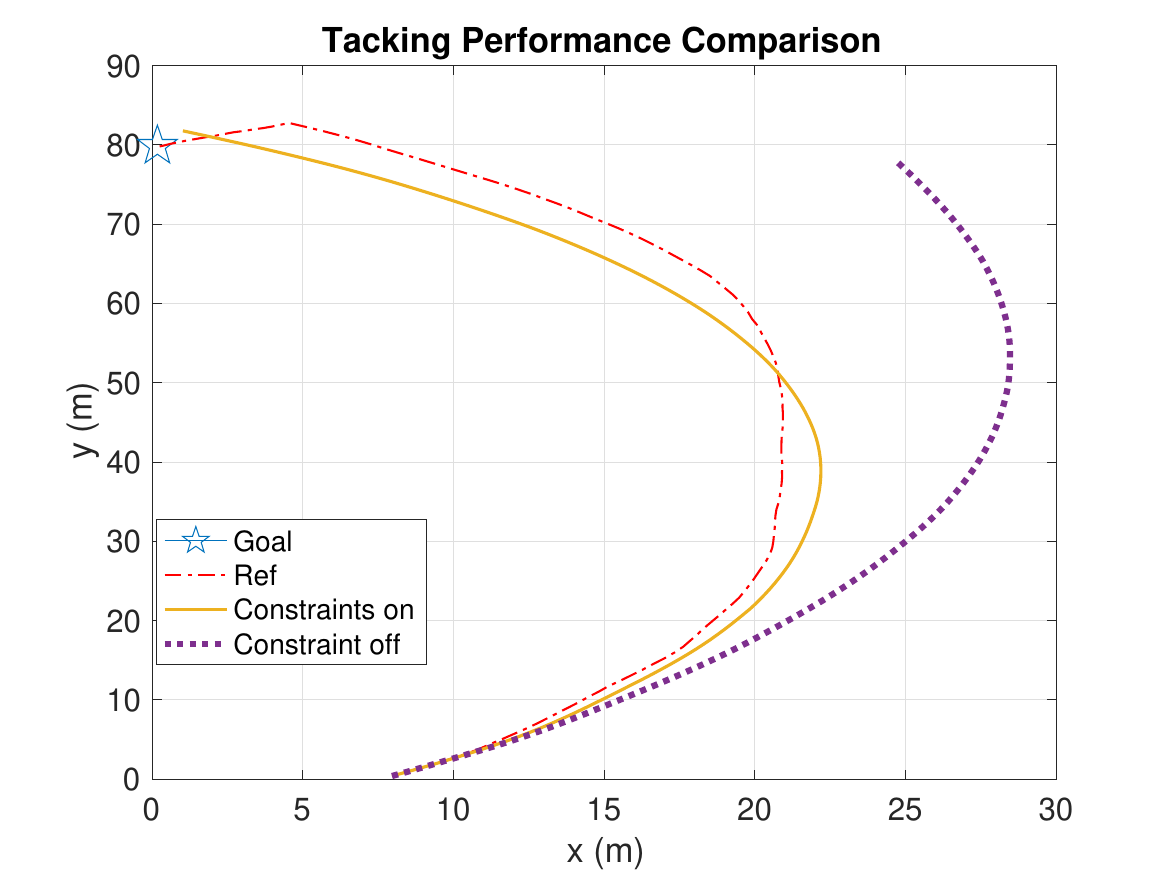}}
    \hfill
	\subfloat[Controllability comparison\label{c5_cons_plot}]{\includegraphics[width=0.49\linewidth]{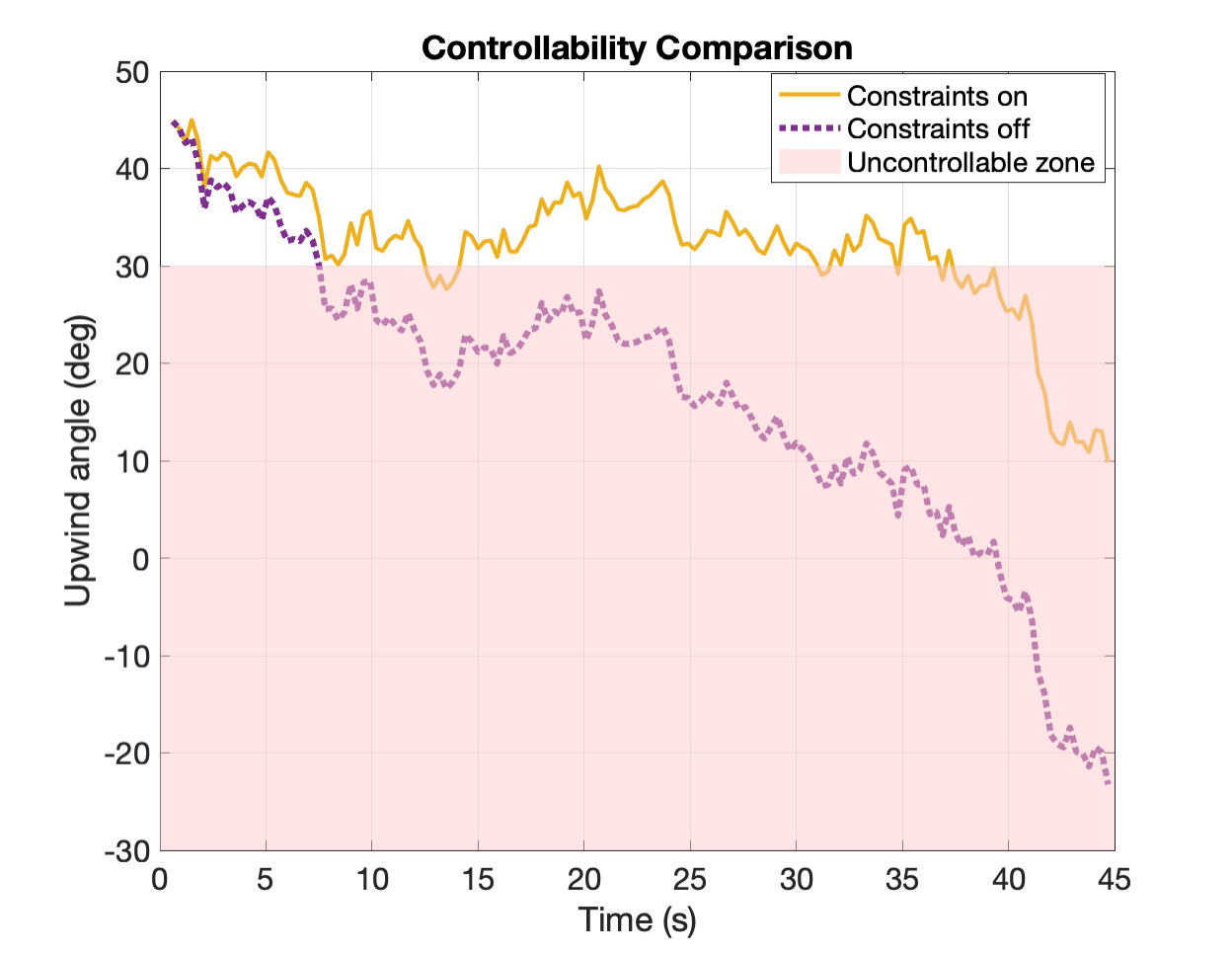}}
	\caption{Comparison between baseline NMPC and controllability-aware NMPC.}
	\label{c5_cons}
\end{figure*}
Fig.~\ref{c5_cons} demonstrates that enforcing controllability constraints improves both trajectory realization and maneuvering quality. In Fig.~\ref{c5_cons}(a), the controllability-analyzed controller tracks the intended route more closely in stochastic wind. At the end of the simulation, the ``constraints on'' case terminates at point $(1.08~\text{m},\,81.77~\text{m})$, whereas the ``constraints off'' terminates at the coordinates $(24.45~\text{m},\,77.37~\text{m})$. The Euclidean distance to the goal is 2.07~m for the controllability-analyzed case, compared to 24.59~m for the baseline case, indicating an improvement by an order of magnitude in final positioning accuracy.

Fig.~\ref{c5_cons}(b) further illustrates the controllability benefit. The shaded region denotes a low-controllability zone, which is approximately within $\pm 30^\circ$ upwind. The controller analyzed for controllability maintains more favourable upwind angles throughout most of the maneuver and avoids prolonged occupancy of the low-controllability region observed in the baseline solution. Near the end of the simulation, the controllability-analyzed solution gradually approaches the low-controllability zone, indicating a trade-off between the tracking accuracy and strict avoidance. In this case, the optimizer favours faster convergence to the target due to the higher weight on the tracking objective. Overall, these results confirm that incorporating controllability analysis into the planning and control loop significantly improves the robustness in dynamic and uncertain wind environments.

\subsection{Discussion and Limitations}
Two limitations of this study should be noted. First, the stochastic wind model is exploited by the planner, whereas the NMPC does not model the wind uncertainty over its prediction horizon and relies on feedback to reject it. The hedging performed by the planner is therefore not coordinated with the controller, which may yield over-conservative reference trajectories, while the controller itself does not adapt its maneuverability margins to the wind uncertainty, as illustrated by the trade-off observed near the end of the maneuver in Fig.~\ref{c5_cons}(b). Propagating the wind statistics into the NMPC, e.g., through scenario-based, tube-based, or chance-constrained formulations, would address both aspects. Second, the evaluation is limited to simulations and to a comparison between the NMPC with and without the proposed constraints. This isolates the effect of the constraints, but comparisons with other guidance laws such as LoS-based controllers, Monte Carlo studies over wind realizations, and experimental validation are needed to assess the approach more broadly.

\section{Conclusion and Future Work}
This paper presented an integrated planning and control framework for autonomous sailing in stochastic wind environments. In contrast to conventional approaches that rely on deterministic wind assumptions, the proposed method explicitly accounts for realistic wind variability, including random gusts and directional fluctuations. A Lie-algebraic analysis of the sailboat dynamics was used to identify the states at which first-order control authority in surge is lost, and the resulting conditions were embedded as constraints in the NMPC. Simulation results indicate that the proposed framework significantly improves the practical feasibility of planned trajectories in the presence of stochastic disturbances. By incorporating system dynamics and constraints within the control architecture, the proposed approach generates efficient and dynamically realizable paths while keeping the vessel away from maneuvering-degraded conditions.
Future work will focus on extending the framework towards real-world deployment. This includes further experimental validation on a physical sailboat platform and investigation of stochastic or chance-constrained MPC formulations that propagate the wind uncertainty over the prediction horizon, as well as comparisons with additional guidance baselines. 
\bibliographystyle{IEEEtran}
\bibliography{Mybibfile}

\begin{thebibliography}{10}
\providecommand{\url}[1]{#1}
\csname url@samestyle\endcsname
\providecommand{\newblock}{\relax}
\providecommand{\bibinfo}[2]{#2}
\providecommand{\BIBentrySTDinterwordspacing}{\spaceskip=0pt\relax}
\providecommand{\BIBentryALTinterwordstretchfactor}{4}
\providecommand{\BIBentryALTinterwordspacing}{\spaceskip=\fontdimen2\font plus
\BIBentryALTinterwordstretchfactor\fontdimen3\font minus
  \fontdimen4\font\relax}
\providecommand{\BIBforeignlanguage}[2]{{%
\expandafter\ifx\csname l@#1\endcsname\relax
\typeout{** WARNING: IEEEtran.bst: No hyphenation pattern has been}%
\typeout{** loaded for the language `#1'. Using the pattern for}%
\typeout{** the default language instead.}%
\else
\language=\csname l@#1\endcsname
\fi
#2}}
\providecommand{\BIBdecl}{\relax}
\BIBdecl

\bibitem{Ferretti}
\BIBentryALTinterwordspacing
R.~Ferretti and A.~Festa, ``Optimal route planning for sailing boats: A hybrid
  formulation,'' \emph{Journal of Optimization Theory and Applications}, vol.
  181, no.~3, pp. 1015--1032, 2019. [Online]. Available:
  \url{https://doi.org/10.1007/s10957-019-01506-x}
\BIBentrySTDinterwordspacing

\bibitem{sto}
C.~Miles and A.~Vladimirsky, ``Stochastic optimal control of a sailboat,''
  \emph{IEEE Control Systems Letters}, pp. 2048--2053, 2022.

\bibitem{jmse11020460}
S.~Liu, Z.~Yu, T.~Wang, Y.~Chen, Y.~Zhang, and Y.~Cai, ``{MPC}-based
  collaborative control of sail and rudder for unmanned sailboat,''
  \emph{Journal of Marine Science and Engineering}, vol.~11, no.~2, 2023.

\bibitem{TIPSUWAN2023114879}
Y.~Tipsuwan, P.~Sanposh, and N.~Techajaroonjit, ``Overview and control
  strategies of autonomous sailboats---a survey,'' \emph{Ocean Engineering},
  vol. 281, p. 114879, 2023.

\bibitem{shen}
C.~Shen, Y.~Shi, and B.~Buckham, ``Integrated path planning and tracking
  control of an {AUV}: A unified receding horizon optimization approach,''
  \emph{IEEE/ASME Transactions on Mechatronics}, vol.~22, no.~3, pp.
  1163--1173, 2017.

\bibitem{Abdelaal}
M.~Abdelaal, M.~Franzle, and A.~Hahn, ``Nonlinear model predictive control for
  trajectory tracking and collision avoidance of underactuated vessels with
  disturbances,'' \emph{Ocean Engineering}, vol. 160, 07 2018.

\bibitem{8126875}
C.~Shen, Y.~Shi, and B.~Buckham, ``Trajectory tracking control of an autonomous
  underwater vehicle using lyapunov-based model predictive control,''
  \emph{IEEE Transactions on Industrial Electronics}, vol.~65, no.~7, pp.
  5796--5805, 2018.

\bibitem{10640017}
H.~Gu and C.~Shen, ``Fast {NMPC} design for image-based visual servoing of
  autonomous underwater vehicles,'' in \emph{2024 IEEE 7th International
  Conference on Industrial Cyber-Physical Systems (ICPS)}, 2024, pp. 1--6.

\bibitem{7799190}
C.~Shen, Y.~Shi, and B.~Buckham, ``Nonlinear model predictive control for
  trajectory tracking of an {AUV}: A distributed implementation,'' in
  \emph{2016 IEEE 55th Conference on Decision and Control (CDC)}, 2016, pp.
  5998--6003.

\bibitem{9816891}
C.~Shen and Y.~Shi, ``{NMPC} design for {AUV} dynamic positioning control with
  incremental input constraints,'' in \emph{2022 IEEE 5th International
  Conference on Industrial Cyber-Physical Systems (ICPS)}, 2022, pp. 1--6.

\bibitem{shen2023marinebook}
Y.~Shi, C.~Shen, H.~Wei, and K.~Zhang, \emph{Advanced Model Predictive Control
  for Autonomous Marine Vehicles}.\hskip 1em plus 0.5em minus 0.4em\relax
  Springer, 2023.

\bibitem{boscain2019}
U.~Boscain and M.~Sigalotti, ``Introduction to controllability of non-linear
  systems,'' in \emph{Contemporary Research in Elliptic {PDEs} and Related
  Topics}.\hskip 1em plus 0.5em minus 0.4em\relax Springer, 2019.

\bibitem{sussmann1972}
H.~J. Sussmann and V.~Jurdjevic, ``Controllability of nonlinear systems,''
  \emph{Journal of Differential Equations}, vol.~12, no.~1, pp. 95--116, 1972.

\bibitem{xiao}
L.~Xiao and J.~Jouffroy, ``Modeling and nonlinear heading control of sailing
  yachts,'' \emph{IEEE Journal of Oceanic Engineering}, vol.~39, no.~2, pp.
  256--268, 2014.

\end{thebibliography}
\end{document}